\documentclass[reprint, superscriptaddress,amsmath,amssymb,aps,prx]{revtex4-2}
\usepackage{graphicx}
\usepackage{dcolumn}
\usepackage{bm}
\usepackage{caption}
\usepackage{subcaption}
\usepackage{gensymb}
\usepackage{xcolor}

\graphicspath{ {./Images/} }
\begin{document}

\title{A route towards stable homochiral topological textures in A-type antiferromagnets}
\date{\today}

\author{Jack Harrison}
\affiliation{Clarendon Laboratory, Department of Physics, University of Oxford, Oxford, OX1 3PU, UK}
\author{Hariom Jani}
\affiliation{Department of Physics, National University of Singapore, 117411, Singapore}
\author{Paolo G. Radaelli}
\email{p.g.radaelli@physics.ox.ac.uk}
\affiliation{Clarendon Laboratory, Department of Physics, University of Oxford, Oxford, OX1 3PU, UK}

\begin{abstract}
Topologically protected whirling magnetic textures could emerge as data carriers in next-generation post-Moore computing. Such textures are abundantly observed in ferromagnets (FMs); however, their antiferromagnetic (AFM) counterparts are expected to be even more relevant for device applications, as they promise ultra-fast, deflection-free dynamics whilst being robust against external fields. Unfortunately, they have remained elusive, hence identifying materials hosting such textures is key to developing this technology. Here, we present comprehensive micromagnetic and analytical models investigating topological textures in the broad material class of A-type antiferromagnets, specifically focusing on the prototypical case of $\alpha \text{-Fe}_2 \text{O}_3$ --- an emerging candidate for AFM spintronics. By exploiting a symmetry breaking interfacial Dzyaloshinskii-Moriya interaction (iDMI), it is possible to stabilize a wide topological family, including AFM (anti)merons and bimerons and the hitherto undiscovered AFM skyrmions. Whilst iDMI enforces homochirality and improves the stability of these textures, the widely tunable anisotropy and exchange interactions enable unprecedented control of their core dimensions. We then present a unifying framework to model the scaling of texture sizes based on a simple dimensional analysis. As the parameters required to host and tune homochiral AFM textures may be obtained by rational materials design of $\alpha \text{-Fe}_2 \text{O}_3$, it could emerge as a promising platform to initiate AFM topological spintronics.
\end{abstract}

\maketitle

\section{ \label{sec:Intro} Introduction}

Topologically protected magnetic textures such as skyrmions and bimerons are emerging as prime information carriers in post-Moore memory and logic devices. Most research has focused on ferromagnetic (FM) materials where examples of such textures are abundant \cite{2020_roadmap, Skyrmappl, SkyrmReview}, but deleterious effects intrinsic to FM topological textures, such as transverse deflection due to the skyrmion Hall effect, preclude their successful integration into devices. As a result, attention has shifted recently to antiferromagnets (AFMs), comprising oppositely aligned magnetic sublattices, as they are predicted to host ultra-small skyrmions that are stable in the absence of applied fields and can be driven at very fast speeds \cite{AFMfunc, AFMSkyrm1, AFMSkyrm2}. However, no examples of isolated AFM skyrmions have, as of yet, been reported in the literature. Recently, there have been promising results in synthetic antiferromagnets \cite{SAFskyrmion1, SAFskyrmion2}. Whilst synthetic AFMs solve the issue of lateral deflection, they may not be able to fully replicate the current-driven `relativistic' physics of natural AFMs due to their weaker interfacial exchange \cite{AFM_DW_SOT}. Hence, discovering such topological textures in natural AFMs remains a key goal of the community.

Topological textures are typically stabilized by an inhomogeneous antisymmetric exchange term called the Dzyaloshinskii-Moriya interaction (DMI), with the materials bulk DMI providing the required energy in many skyrmion-hosting systems \cite{2020_roadmap}. Materials without bulk DMI can still host stable chiral textures if an interfacial DMI (iDMI) is induced at the material surface \cite{iDMIEnergy}, usually via an interaction with an over/under-layer that has strong spin-orbit coupling \cite{iDMIanatomy}. The iDMI tends to favor N\'{e}el type textures of a fixed chirality --- an important feature since spin torque driven motion depends on chirality \cite{SkyrmHallinAFM}. Homochiral topological AFM textures are expected to move consistently and reproducibly under the action of spin currents at speeds of up to a few km\,s$^{-1}$, making them promising as non-volatile information carriers in spintronic devices \cite{AFM_DW_SOT, AFM_SkyrmMotion, BimChaos, SkyrmHallinAFM}. 

Here, we focus on $\alpha \text{-Fe}_2 \text{O}_3$, which is a promising material candidate for AFM spintronics as it exhibits ultra-low Gilbert damping and has exceptionally-long and tunable spin diffusion \cite{SpDiff, SpTrans}, shows a sizable spin-Hall magnetoresistance \cite{SpHall1, SpHall2} and its AFM domain configurations can potentially be switched using pulsed currents through heavy-metal overlayers \cite{FeCurrent1, FeCurrent2}. We previously reported the discovery of flat (anti)vortices in $\alpha \text{-Fe}_2 \text{O}_3$ thin films coupled to a ferromagnetic Co overlayer \cite{Francis} and, more recently, of topological merons, antimerons and bimerons in films with a Pt overlayer \cite{Hariom}. In the latter case, we were able to repeatedly nucleate and destroy these topological textures via thermally cycling through the spin-reorientation `Morin' transition, which is in some ways analogous to a Kibble-Zurek quench \cite{Kibble, Zurek, KZMech}. Since our observed textures were of both Bloch and N\'{e}el types and thus were not homochiral, we deduced that our samples had negligible iDMI. Understanding how to tune both the chirality and scale of AFM topological textures in the presence of iDMI is crucial for spintronics applications such as AFM topological racetracks.

$\alpha \text{-Fe}_2 \text{O}_3$ crystallizes in the corundum structure (space group R$\bar{3}$c) and is an antiferromagnet with a relatively high N\'{e}el temperature ($\approx 960$ K \cite{CoeyBook}). The magnetic moments in the antiferromagnetic phase stack antiparallel along the $c$-axis, such that moments in each $a$-$b$ plane are ferromagnetically coupled \cite{MagStruct}, see Fig. \ref{fig:magstruct}. This spin arrangement is generally known as A-type from the classic field of perovskite magnetism and we will use this terminology herein. $\alpha \text{-Fe}_2 \text{O}_3$ also hosts the Morin transition \cite{Morin} at $T_\text{M} \approx 260$ K in bulk samples, where the anisotropy of the Fe$^{3+}$ ions flips from being $a$-$b$ easy-plane for $T>T_\text{M}$ to easy-axis along the c-axis for $T<T_\text{M}$ due to a competition between on-site and dipolar anisotropies \cite{MorrishBook, BPAnis, FeAnis, HDoping}. The resulting net anisotropy is strongly temperature-dependent and changes sign at $T_\text{M}$. Consequently, both easy-plane and easy-axis domain morphologies can easily be studied via in-situ temperature variations \cite{Francis, Hariom, HDoping, Anis_SpHall}.

In this paper, we explore the effects of iDMI on a wide family of topological textures in the easy-plane and easy-axis phases of $\alpha \text{-Fe}_2 \text{O}_3$ by analytical calculations and micromagnetic simulations. We find that such textures become homochiral, making them ideal for spintronics applications where they can potentially be moved at ultra-fast speeds via spin-orbit torques \cite{BimChaos}. Moreover, their stability and size can be carefully controlled as a function of the material parameters to achieve the requirements for applications. A key prediction of our paper is that \emph{antiferromagnetic skyrmions} should be stable in this system below the Morin transition for a wide range of physically realistic material parameters, making this an exciting and promising platform for their experimental discovery.

\section{ \label{sec:Model} Micromagnetic model}
\subsection{ \label{sec:General} General approach for A-type antiferromagnets}

\begin{figure}
\centering
\begin{subfigure}[t]{0.18\textwidth}
\caption{}
\centering
\includegraphics[width=\textwidth]{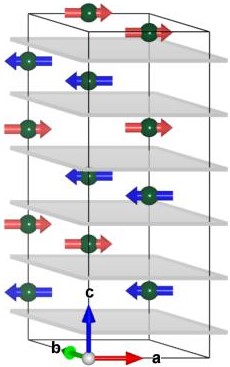}
\label{fig:magstruct}
\end{subfigure}
\hfill
\begin{subfigure}[t]{0.295\textwidth}
\caption{}
\centering
\includegraphics[width=\textwidth]{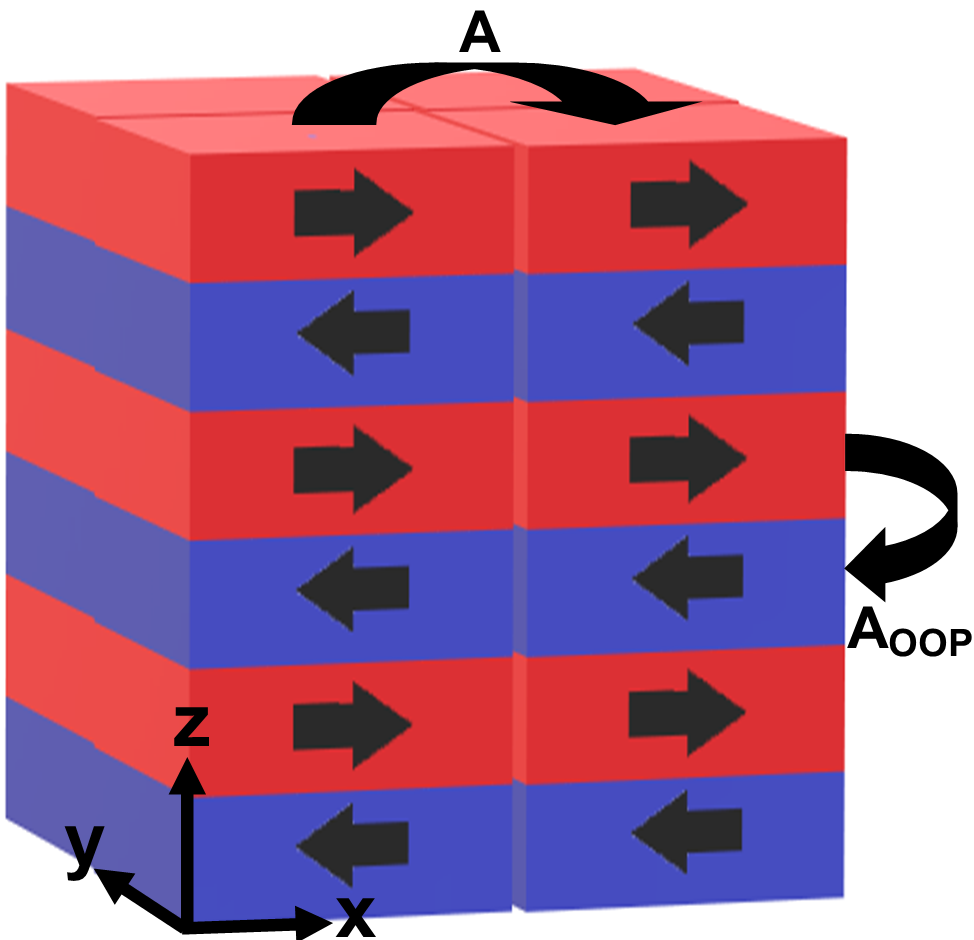}
\label{fig:config}
\end{subfigure}
\caption{a) The crystal and magnetic structures of AFM $\alpha \text{-Fe}_2 \text{O}_3$, showing a sequence of antiparallel FM layers that we model in our simulations. The light red and dark blue arrows show the orientation of the magnetic moments in the two sublattices above $T_\text{M}$. The gray planes separate the two sublattices and host the O-atoms. b) The simulation configuration, where cells belonging to the two antiparallel sublattices are shown as red/blue and the arrows show the magnetic moment orientation in a given cell for $T>T_\text{M}$. The curved arrows show the AFM coupling between adjacent layers ($A_{\text{OOP}}$) and the FM coupling between cells in the same layer ($A$).}
\label{fig:setup}
\end{figure}

The application of micromagnetic modeling techniques and code such as \textsc{mumax3} \cite{mumax1} to ferromagnetic bulk, films and multi layers is extremely well documented. Whilst some applications to antiferromagnets can also be found in the literature \cite{GAFMmodel, MicroAFM}, it is not immediately obvious that these techniques can be applied to all antiferromagnets. Generally, one considers the micromagnetic scale to be much larger than the atomic scale, so that various magnetic interactions (exchange, anisotropy and dipolar) can be replaced by their continuum counterparts. This assumption does not hold for an antiferromagnet, since the magnetization changes sign within every unit cell. In the case of A-type antiferromagnets such as $\alpha \text{-Fe}_2 \text{O}_3$ the situation is somewhat simpler, since these materials consist of ferromagnetic layers stacked antiparallel along an axis. 

Here, we choose to model a generic A-type antiferromagnet as a set of layers stacked along the $z$-axis of our simulation space, Fig. \ref{fig:config}. We consider the magnetic moments to be rigidly coupled along the $z$-axis so that antiferromagnetic alignment is strictly enforced throughout the material. Akin to previous models of synthetic antiferromagnets \cite{SAFskyrmion1, SAFskyrmion2}, here the micromagnetic cell in the $x$-$y$ plane is chosen to be much larger than the lattice parameter $a$, whereas its size along the $z$-axis corresponds to the spacing of the ferromagnetic sublayers (1/6 of the lattice parameter $c$). We will focus throughout on the specific case of $\alpha \text{-Fe}_2 \text{O}_3$; however, this approach is applicable to general A-type antiferromagnets if the relevant material parameters are used. The key energy terms are the exchange, anisotropy, dipolar and interfacial Dzyaloshinskii-Moriya interactions (iDMI), whose forms are given in Appendix \ref{App:Eterms}. 

Our approach differs from previous micromagnetic models of antiferromagnets \cite{MicroAFM, GAFMmodel}, since here the spatial separation of the layers along the $z$-axis has a physical meaning and we treat the dipolar fields organically rather than neglecting them or assuming they can be fully subsumed within the anisotropy. The dipolar fields require special attention as they are typically long-range interactions, whereas all the other terms are short-range interactions between adjacent moments and are subsumed into scale-independent macroscopic parameters (see Appendix \ref{App:param}). In essence, whereas the dipolar interaction decreases very rapidly at macroscopic distances, as expected for an antiferromagnet, its short-range component results in an in-plane (IP) anisotropy, so that the effective anisotropy of an A-type antiferromagnet, $K_{\text{eff}}$ is the the sum of a dipolar component $K_{\text{dip}}$ and of an on-site component $K_{\text{os}}$ (if any exists). This is entirely physical, and the fine balance of $K_{\text{os}}$ and $K_{\text{dip}}$ is indeed the origin of the Morin transition in $\alpha \text{-Fe}_2 \text{O}_3$ \cite{FeAnis}. Whilst this accounts for the largest component of the dipolar fields, there may also be some small effects due to stray fields at the uncompensated surface layers and any such effects will be included in our simulations.

It should be noted that our model has a number of limitations applying to all A-type antiferromagnets. Firstly, as already mentioned, the antiferromagnetic exchange interaction along the $z$-axis ensures that spins in adjacent layers are exactly antiparallel. Therefore, no variations in textures along the $z$-axis can be studied as this model does not accurately reproduce the atomic-scale interactions along this direction. By contrast, our approach is well suited to studying magnetic textures that are modulated in the $x$-$y$ plane. In this work we only consider very thin films, with total thickness much smaller than the characteristic length scale for magnetic variations along the $z$-direction, hence this limitation will have a negligible effect on our simulations. Secondly, the dynamics at the nm length scale, primarily AFM spin waves, cannot be accurately simulated using this model \cite{GAFMmodel}, making it inappropriate for studying dynamical phenomena. Since our focus is the study of steady-state magnetization configurations via energy minimization, this drawback will also not affect our results.

There are a few additional caveats in the specific case of $\alpha \text{-Fe}_2 \text{O}_3$. Firstly, this system has a weak bulk DMI, which causes a small canting of the two sublattices when $T>T_\text{M}$ leading to a small ferromagnetic component \cite{MorrishBook}. We neglect such a bulk DMI and its associated small canting throughout, as analytical calculations suggest that it cannot by itself stabilize IP-modulated topological textures. It should be noted that the small canting will become relevant if we want to extend the model to incorporate the effect of externally applied magnetic fields, which are not discussed here. Additionally, this system has a weak basal plane anisotropy due to its trigonal crystal structure, which favors the formation of a set of three 120\textdegree ~domains and their time reversed counterparts when $T>T_\text{M}$; however, this anisotropy is orders of magnitude weaker than the uniaxial on-site anisotropy \cite{BPAnis} and is therefore neglected here.

\subsection{ \label{sec:Micro} Micromagnetic simulations}

\begin{figure*}
\centering
\begin{subfigure}[t]{0.232\textwidth}
\caption{N\'{e}el meron}
\centering
\includegraphics[width=\textwidth]{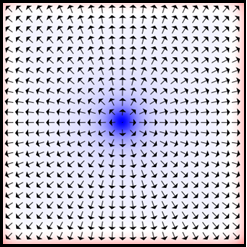}
\label{fig:Meron}
\end{subfigure}
\hfill
\begin{subfigure}[t]{0.231\textwidth}
\caption{Distorted antimeron}
\centering
\includegraphics[width=\textwidth]{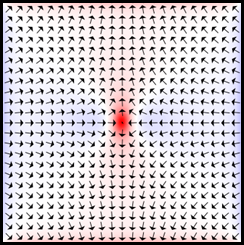}
\label{fig:AMeron}
\end{subfigure}
\hfill
\begin{subfigure}[t]{0.234\textwidth}
\caption{N\'{e}el Bimeron}
\centering
\includegraphics[width=\textwidth]{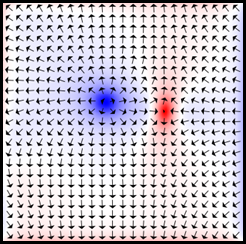}
\label{fig:Bimeron}
\end{subfigure}
\hfill
\begin{subfigure}[t]{0.282\textwidth}
\caption{N\'{e}el Skyrmion}
\centering
\includegraphics[width=\textwidth]{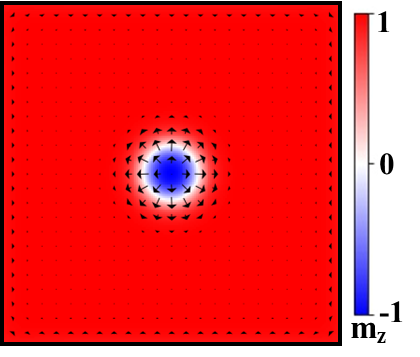}
\label{fig:Skyrmion}
\end{subfigure}
\caption{A gallery of topological textures in $\alpha \text{-Fe}_2 \text{O}_3$, based on our micromagnetic simulations. In all cases only a single magnetic layer is shown. The black arrows represent the in-plane spin directions and the color contrast corresponds to the $z$-component of magnetization.}
\label{fig:Textures}
\end{figure*}

Our simulations were performed using \textsc{mumax3} \cite{mumax1, mumax2, mumax3}, an open-source micromagnetics package utilizing finite-element simulations to model magnetic structures on the nanometer to micrometer length scales. This is ideal for topological textures, which tend to be around 100 nm or smaller \cite{Hariom, 2020_roadmap}. The micromagnetic solver compares several energy terms as discussed in section \ref{sec:General}. For each simulation, the system was initialized with a certain configuration (meron, skyrmion, etc.), which was allowed to evolve using the conjugate gradient method \cite{mumax2} to minimize its energy. Given that we are considering \emph{topological} structures, which are generally protected from collapsing due to a finite energy barrier, the minimization procedure should only alter their size and geometry in a manner determined by the competition of the relevant energy terms, independently of the exact procedure used; for example, evolving the texture dynamically using the Landau-Lifshitz-Gilbert equation with large damping \cite{AmikamBook} should result in the same final topology. A texture whose final topology (after energy minimization) is identical to the initial topology is considered `stable' in these simulations, though it is generally metastable with respect to a uniform spin configuration. Minimized configurations for the key topological textures studied here can be found in Fig. \ref{fig:Textures}.

The simulation is split into cuboid cells, each assigned a magnetic moment corresponding to the sublattice magnetization that is constant in each cell. Full details of the simulation parameters used are given in Appendix \ref{App:param}. The demagnetizing field is calculated automatically in the software by convolving the magnetization field with the demagnetizing kernel \cite{mumax1}, resulting in a demagnetization contribution to the total energy. In all our simulations this results in an effective easy-plane anisotropy of strength $K_{\text{dip}} \approx 530$ kJ\,m$^{-3}$ perpendicular to the $z$-axis, consistent with our discussion of the dipolar fields in section \ref{sec:General}. By applying an additional uniaxial on-site anisotropy of strength $K_\text{os}$ along the $z$-axis, we can simulate a Morin-like transition by varying $K_\text{os}$ around $K_{\text{dip}}$. The effective anisotropy constant is given by $K_{\text{eff}} = K_\text{os}- K_{\text{dip}}$ and can switch sign from positive to negative, corresponding to out-of-plane (OOP) or IP orientations below or above the Morin transition, respectively. Performing a set of simulations without the demagnetizing fields and a readjusted anisotropy constant instead ($K_{\text{dip}}=0, K_{\text{eff}}= K_\text{os}$) resulted in identical scaling of topological textures (see Appendix \ref{App:NoDMI}). This suggests that any other effects of the demagnetizing fields are negligible when studying the static properties of topological textures. In our simulations, we also apply an iDMI of strength $D$ (in the range $\approx 0.5-3$ mJ\,m$^{-2}$) only to the topmost layer, thereby simulating a symmetry broken magnetic surface hosting an interfacial antisymmetric exchange \cite{iDMIanatomy}.

\section{Analytical calculations} \label{sec:analytics}

Extensive analytical work has been performed to understand the shape and size scaling of skyrmions \cite{SkyrmModel, SkyrmSize, AFMSkyrm1, AFMSkyrm2}. Other topological textures, such as merons \cite{GaoMerons, AFMVortex, CrClMeron} and bimerons \cite{GobelBim, AsymmSkyrm, SkvsBim, BimDefects, BimChaos} have also been receiving attention recently. Complicated winding textures such as these are difficult to study analytically, especially when their functional forms are not known exactly. One possible simplification is to impose a particular functional form for the texture, usually called an `ansatz'. This often makes the problem solvable analytically, yielding an exact scaling that can be usefully compared with more realistic simulations. We calculated the exchange, anisotropy and iDMI energy for a linear (anti)meron ansatz (see Appendix \ref{App:Analytics}) in order to derive the associated texture sizes and directly compare them with our micromagnetic simulations. As discussed above, the dipolar field acts as an effective IP anisotropy to first order, therefore this was not included directly in our analytical calculations but rather rolled into the anisotropy.

Our approach here is analogous to calculations we have performed previously \cite{Jacopo, Hariom} but with the addition of the iDMI energy \cite{iDMIEnergy}. For \emph{merons} the effect of the iDMI is relatively straightforward, since it tends to stabilize circular homochiral textures of the N\'{e}el type. Using a linear meron ansatz, the analytical expression for the meron radius $R$ is (Appendix \ref{App:Meron})
\begin{equation}
R = l_w \left(\kappa + \sqrt{\kappa^2 + 1}\right) = \frac{3}{4}F,
\label{eqn:meron}
\end{equation}
where the final term in the above equation relates the meron `radius' $R$, which is the only free parameter of this specific ansatz, to the full-width at half maximum (FWHM) $F$ of the texture, which can be determined from the simulations. In eq. \ref{eqn:meron}, the characteristic length scale is $l_w = \eta \sqrt{A/|K_\text{eff}|}$ and we have introduced the dimensionless parameter $\kappa = \kappa_0 D_{\text{eff}}/\sqrt{A|K_\text{eff}|}$, which describes how strongly the iDMI energy affects the textures relative to the exchange and anisotropy. $D_{\text{eff}} = D/N$ is the rescaled DMI parameter where N is the total number of layers in our model system. Clearly, $R \rightarrow l_w $ in the limit $D_{\text{eff}} \rightarrow 0$. $\eta$ and $\kappa_0$ are ansatz-dependent numerical constants; their values for a linear meron are derived in Appendix \ref{App:Meron}.

By contrast, the situation for \emph{antimerons} is more complex, since they are composed of sectors of alternating chirality. In the presence of iDMI the energetically favored N\'{e}el sectors contract, as the iDMI energy prefers tight spirals, resulting in an elongated (elliptical) antimeron. There are two key parameters that can be extracted by minimizing the antimeron energy (Appendix \ref{App:AntiMeron}), namely the radius $R$ and distortion parameter $\lambda$, giving
\begin{widetext}
\begin{eqnarray}
R &=& l_w\left[\frac{1}{2}\kappa \left(\lambda - \frac{1}{\lambda}\right) + \sqrt{\frac{1}{4}\kappa^2 \left(\lambda - \frac{1}{\lambda}\right)^2 + \frac{1}{2}\left(\lambda^2+\frac{1}{\lambda^2}\right)}\right] = \frac{3}{4}\sqrt{F_{long}F_{short}}, \label{eqn:AMeron1} \\
\lambda &=& \frac{\kappa R}{2l_w\left[C+ ln \left(\frac{R_\text{d}}{R}\right) \right]} + \sqrt{1 + \left\{\frac{\kappa R}{2l_w\left[C+ ln \left(\frac{R_\text{d}}{R}\right) \right]}\right\}^2} = \sqrt{\frac{F_{long}}{F_{short}}},
\label{eqn:AMeron2}
\end{eqnarray}
\end{widetext}
where $l_w$, $\kappa$, $\eta$ and $\kappa_0$ are defined in the same way as for a meron (see Appendix \ref{App:AntiMeron}). $C \approx 2$ is a numerical integration constant. Here, $F_{long}$ and $F_{short}$ correspond to the FWHM along the long and short axes of the antimeron respectively. In our calculation, we confined the antimeron to a region of radius $R_\text{d}$, representing a cut-off on the effect of the antimeron distortion, which would otherwise extend to infinity as a consequence of the analytical approach. These equations have an exact, albeit complicated, solution for $R$ and $\lambda$ for general $\kappa$ and $l_w$ if $R_\text{d}$ is given, therefore we solve Eq's. \ref{eqn:AMeron1} and \ref{eqn:AMeron2} iteratively for given $A$, $K_\text{eff}$ and $D$. The results are convergent for all parameter values relevant here, given a reasonable initial guess of $\lambda$ (see section \ref{sec:AMeron} and Appendix \ref{App:AntiMeron}).

\section{ \label{sec:Simple} Simple easy-plane topological textures}
\subsection{Merons}

\begin{figure*}
\centering
\begin{subfigure}[t]{0.32\textwidth}
\caption{}
\centering
\includegraphics[width=\textwidth]{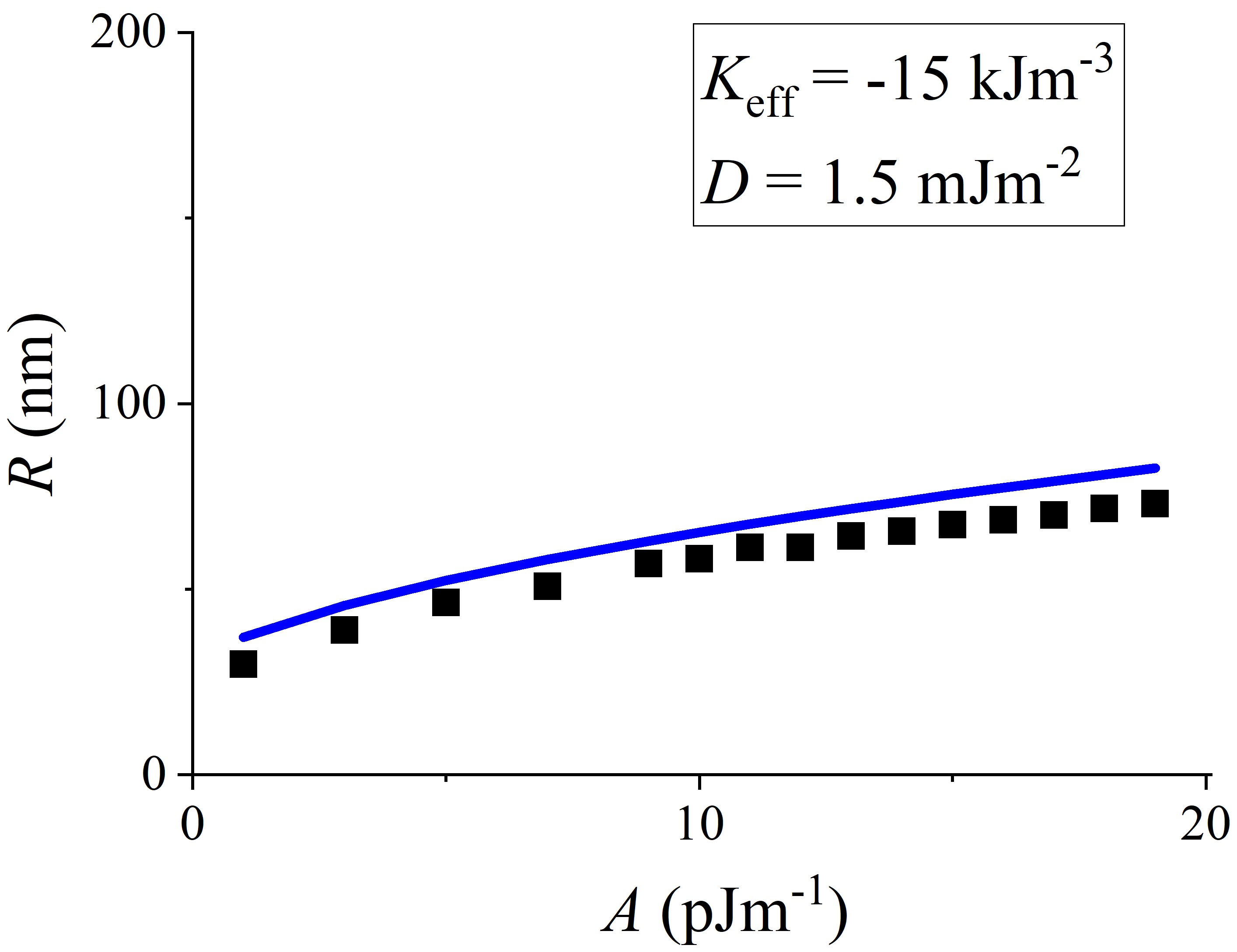}
\end{subfigure}
\hfill
\begin{subfigure}[t]{0.32\textwidth}
\caption{}
\centering
\includegraphics[width=\textwidth]{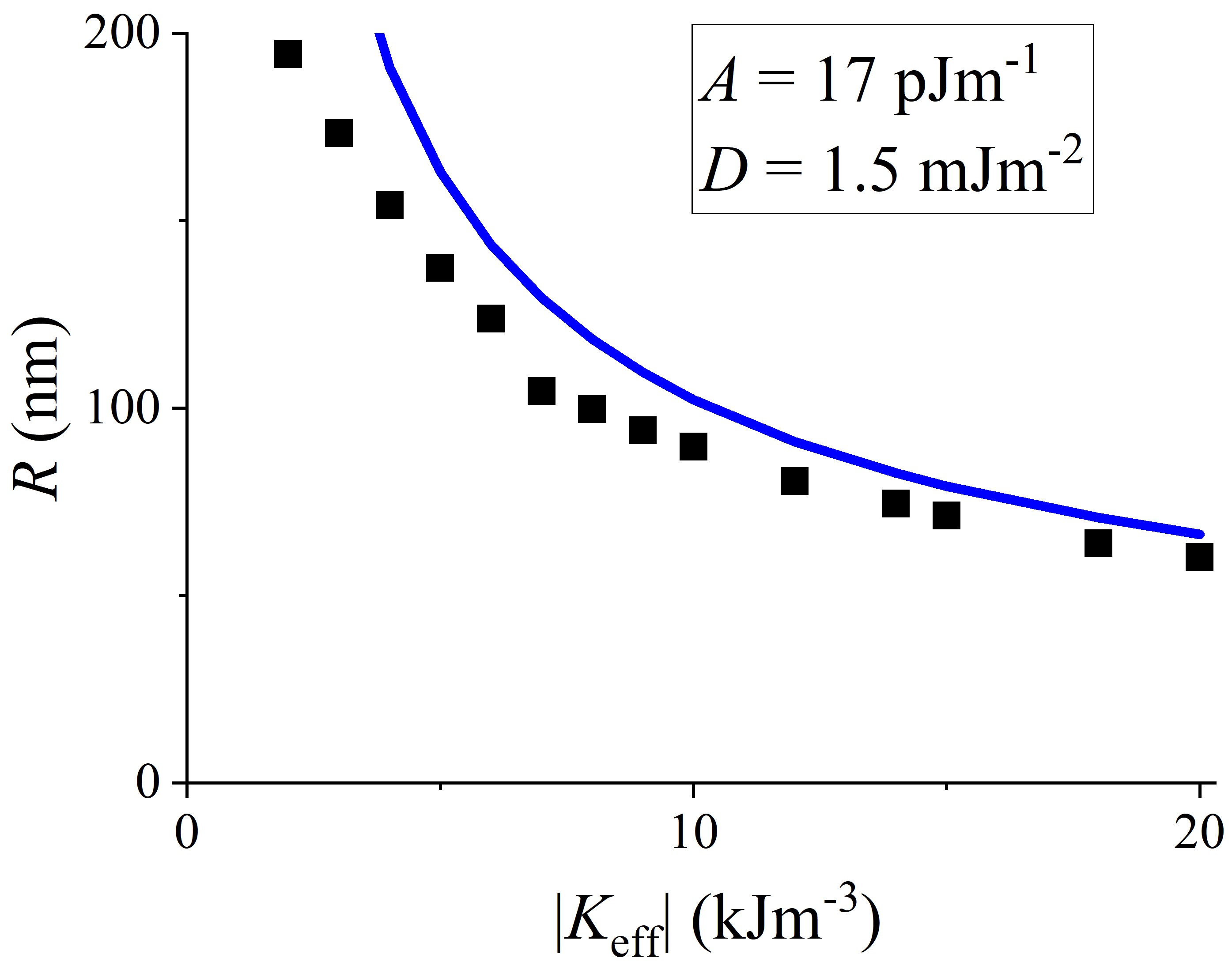}
\end{subfigure}
\hfill
\begin{subfigure}[t]{0.32\textwidth}
\caption{}
\centering
\includegraphics[width=\textwidth]{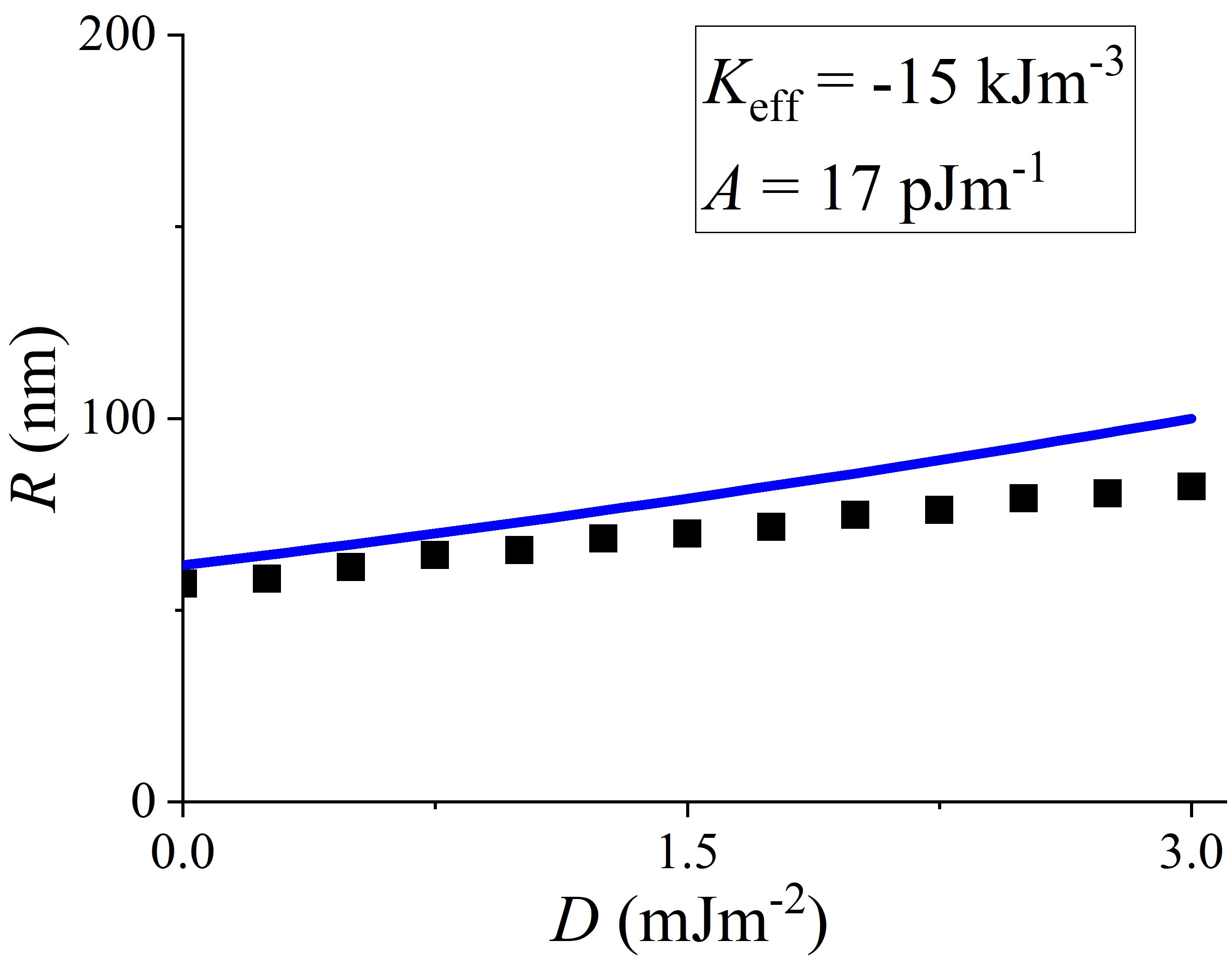}
\end{subfigure}
\caption{Radius $R$ of \emph{meron textures}, based on micromagnetic simulations (black points) and analytical calculations (blue lines). The iDMI was included in all cases. $A$, $|K_\text{eff}|$ and $D$ were varied in panels (a), (b) and (c) respectively, with the rest of the parameters kept constant.}
\label{fig:DMIMeron}
\end{figure*}

As merons require easy-plane anisotropy and therefore are observed in $\alpha \text{-Fe}_2 \text{O}_3$ for $T>T_\text{M}$ \cite{Hariom}, we use values of $K_{\text{os}}$ such that $K_{\text{eff}}<0$. We performed micromagnetic simulations of isolated merons using our model for the case of zero iDMI as a consistency check (see Appendix \ref{App:NoDMI}). To study the effects of iDMI we also performed a set of meron simulations with non-zero $D$ and compared their sizes to the analytical expression in Eq. \ref{eqn:meron}, Fig. \ref{fig:Meron}. As expected, the presence of iDMI enforces a specific chirality, making all such merons N\'{e}el type. The scaling with $A$ and $|K_\text{eff}|$ is, to lowest order, similar to that found for the case of a meron without iDMI \cite{Hariom} and the functional form that we determined analytically provides a satisfactory approximation to the simulations. The key trends are that the meron radius increases if we increase the strength of the exchange or DMI, whilst decreasing rapidly as we increase the strength of the IP anisotropy. There appears to be some difference in the actual FWHM values when comparing the simulations and analytics, which is not surprising, since the numerical prefactors contained in $l_w$ and $\kappa$ are strongly affected by the choice of the ansatz; calculations for a different ansatz would give different numerical factors \cite{Jacopo}. The qualitative agreement between our computational model and analytical calculations demonstrate that our approach is both reasonable and internally consistent.

\subsection{Antimerons} \label{sec:AMeron}

\begin{figure*}
\centering
\begin{subfigure}[lt]{0.32\textwidth}
\caption{}
\centering
\includegraphics[width=\textwidth]{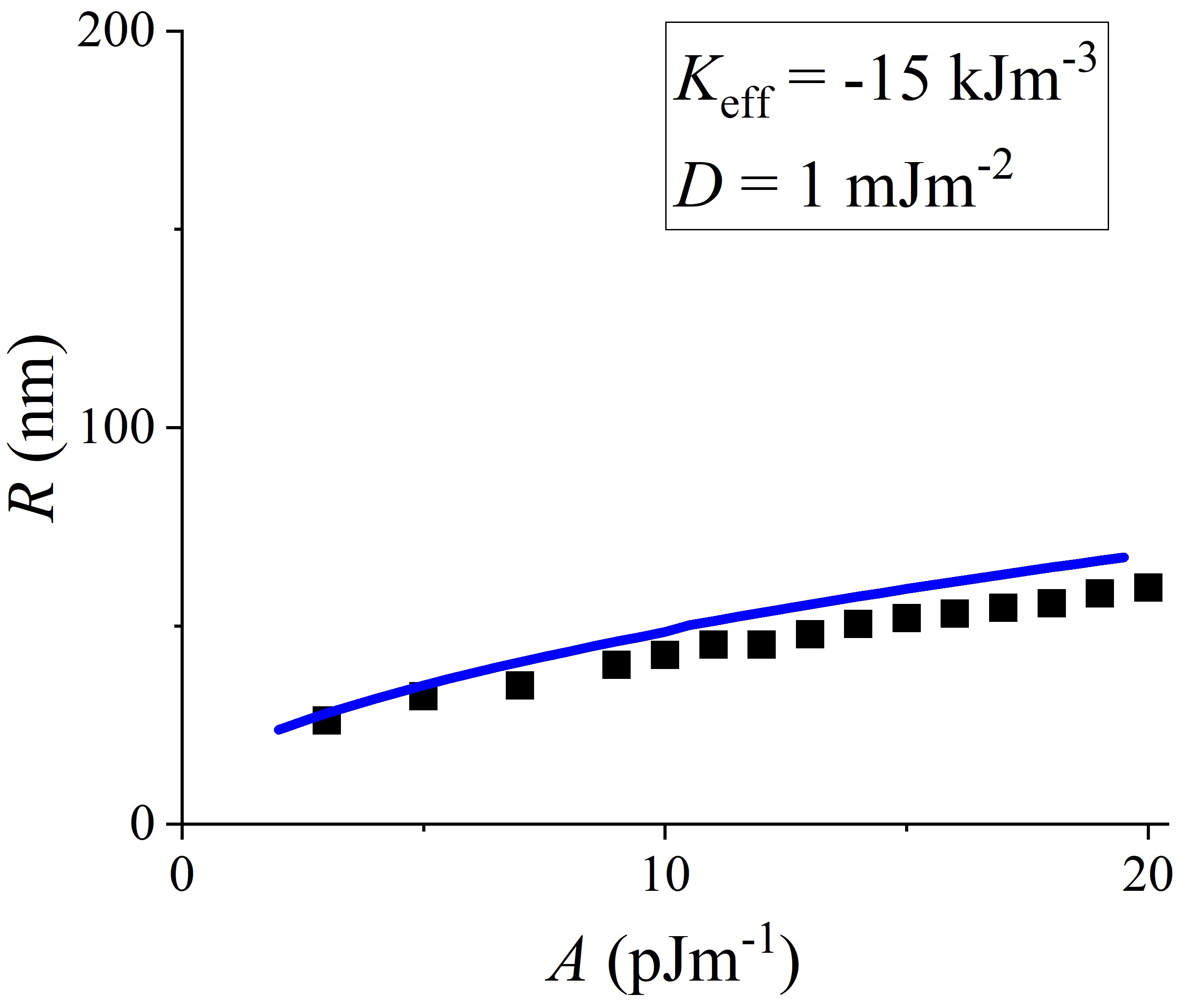}
\end{subfigure}
\hfill
\begin{subfigure}[ct]{0.32\textwidth}
\caption{}
\centering
\includegraphics[width=\textwidth]{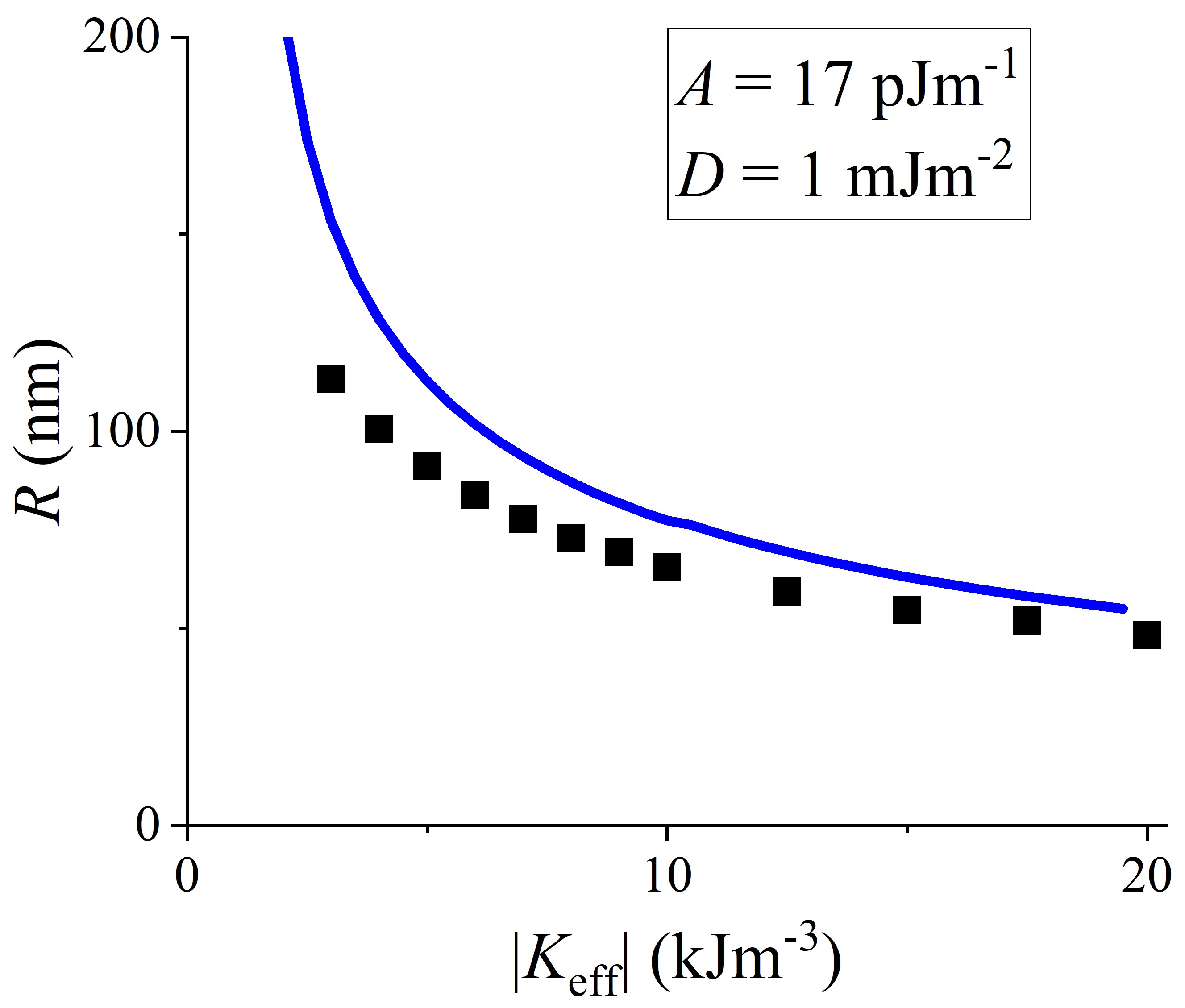}
\end{subfigure}
\hfill
\begin{subfigure}[rt]{0.32\textwidth}
\caption{}
\centering
\includegraphics[width=\textwidth]{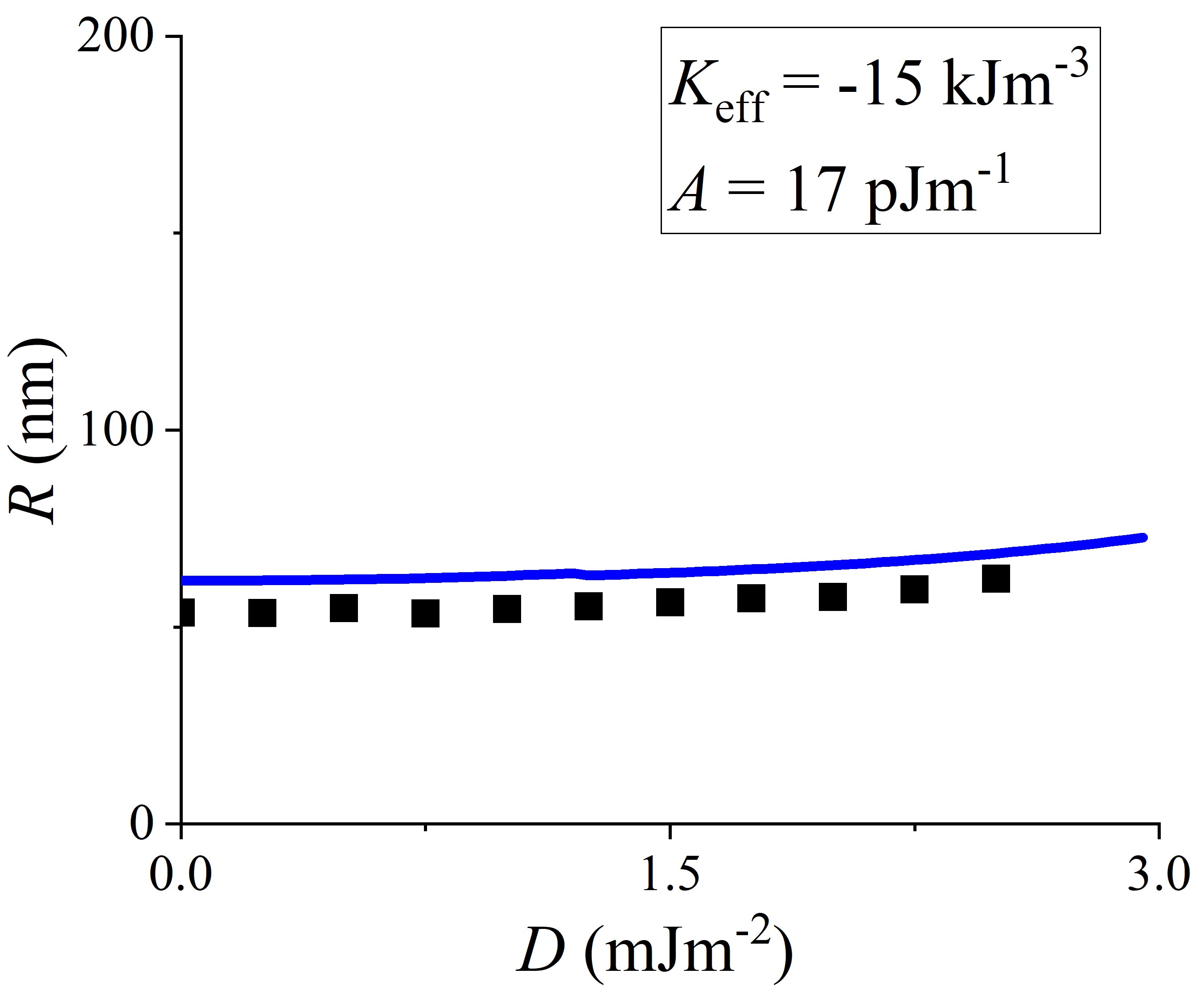}
\end{subfigure}
\caption{Radius $R$ of \emph{antimeron textures}, based on micromagnetic simulations (black points) and analytical calculations (blue lines). The iDMI was included in all cases. $A$, $|K_\text{eff}|$ and $D$ were varied in panels (a), (b) and (c) respectively, with the rest of the parameters kept constant.}
\label{fig:AntiMeron}
\end{figure*}

We performed simulations of isolated antimerons with iDMI and found that they were indeed stable and distorted, Fig. \ref{fig:AMeron}. We have therefore calculated the effective radius $R$ of the simulated antimeron using Eq. \ref{eqn:AMeron1}, and compared it with the analytical values (Fig. \ref{fig:AntiMeron}). The analytical curve was calculated iteratively using Eq's. \ref{eqn:AMeron1} and \ref{eqn:AMeron2} for the same set of values of $A$, $|K_\text{eff}|$ and $D$ used in the simulations and with the cut-off radius $R_\text{d}$ set to the simulation radius. It should be noted that varying $R_\text{d}$, even by an order of magnitude, has a minimal effect on the resulting analytical radius. 

Here, the antimeron radius increases with increasing exchange strength, is roughly independent of the iDMI strength and decreases rapidly as the IP anisotropy increases. The scaling of antimerons as a function of $A$, $|K_\text{eff}|$ and $D$ is qualitatively similar to merons; however, there is again a slight difference between the analytical and simulated radius due to the ansatz choice. Whilst the scaling of the distortion with the various energy terms matches qualitatively, the analytically calculated value of the distortion parameter $\lambda$ is consistently smaller compared to the value extracted from the simulations, by a factor of $\sim3-5$, see Appendix \ref{App:DistGraphs}. Despite these caveats, the scaling behavior of $R$ in both the analytical and simulated antimerons match reasonably well and they both predict that antimerons should distort in the presence of iDMI.

\section{\label{sec:bimerons} Compound easy-plane topological textures}
\subsection{Bimerons and topologically trivial meron pairs}

\begin{figure*}
\centering
\begin{subfigure}{0.55\textwidth}
\caption{}
\centering
\includegraphics[width=\linewidth]{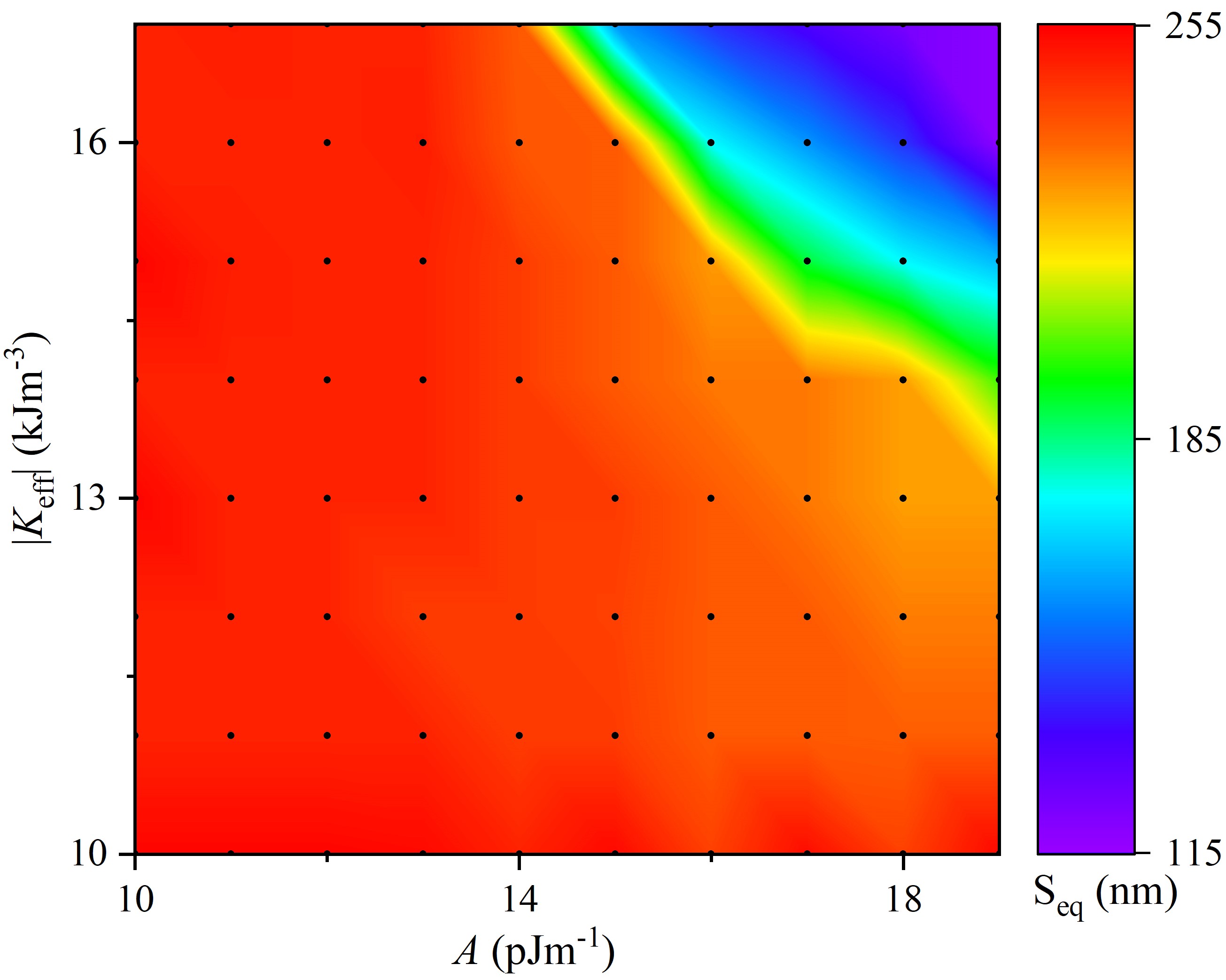}
\end{subfigure}
\hfill
\begin{minipage}{0.4\textwidth}
\begin{subfigure}{0.47\textwidth}
\caption{}
\centering
\includegraphics[width=\linewidth]{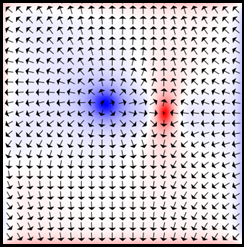}
\end{subfigure}
\hfill
\begin{subfigure}{0.47\textwidth}
\caption{}
\centering
\includegraphics[width=\linewidth]{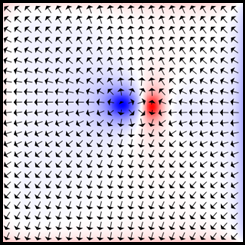}
\end{subfigure}
\\[\baselineskip]
\begin{subfigure}{0.47\textwidth}
\caption{}
\centering
\includegraphics[width=\linewidth]{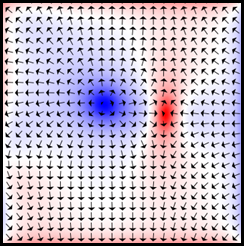}
\end{subfigure}
\hfill
\begin{subfigure}{0.47\textwidth}
\caption{}
\centering
\includegraphics[width=\linewidth]{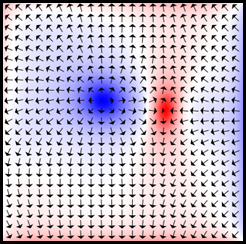}
\end{subfigure}
\end{minipage}

\caption{(a) Relaxed N\'{e}el bimeron inter-core distance $s_\text{eq}$ as a function of $A$ and $|K_\text{eff}|$ for $D=2$ mJ\,m$^{-2}$. The simulation points are given by black squares. (b)-(d) Snapshots of the relaxed configuration in one of the layers for several points in the diagram, where the arrows represent the IP spin direction and the color contrast represents the $z$-component of the magnetization as in Fig. \ref{fig:Textures}. The images correspond to the values: (b) $A=14$ pJ\,m$^{-1}$, $K_\text{eff}=-17$ kJ\,m$^{-3}$, (c) $A=19$ pJ\,m$^{-1}$, $K_\text{eff}=-17$ kJ\,m$^{-3}$, (d) $A=14$ pJ\,m$^{-1}$, $K_\text{eff}=-10$ kJ\,m$^{-3}$, (e) $A=19$ pJ\,m$^{-1}$, $K_\text{eff}=-10$ kJ\,m$^{-3}$.}
\label{fig:Bimscale}
\end{figure*}

In our previous study \cite{Hariom} we reported the observation of meron-antimeron pairs in $\alpha \text{-Fe}_2 \text{O}_3$, which could either be topologically trivial meron pairs (TTMPs) or topologically non-trivial bimerons, depending on whether the core polarization of the constituent (anti)merons are aligned or anti-aligned respectively, Fig. \ref{fig:Bimeron}. Since constructing an analytical model for such compound objects using a realistic ansatz is difficult, we investigated their properties using micromagnetic simulations. We initialize a meron-antimeron pair in the system that is either a TTMP or a bimeron in the IP state ($K_{\text{eff}}<0$) by placing a meron in one half of the simulation and an antimeron in the other half with the desired core polarities. We then allow them to relax naturally into their preferred configuration. We observe that neither TTMPs nor bimerons are stable in the absence of iDMI. The observed collapse indicates that the competition between exchange and anisotropy energies alone is insufficient to stabilize such textures, despite the supposed topological protection of bimerons.

If we introduce iDMI into the system, bimerons become stable over a very wide parameter range, even when initialized at very close distances ($\leq 150$ nm), Fig. \ref{fig:Bimscale}, whereas TTMPs are only stable if they start a long way apart, at which point they could be considered as isolated (anti)merons. This makes phenomenological sense as bimerons have a net topological charge and therefore should be prevented from collapsing due to a finite energy barrier afforded by iDMI, which is expected to be absent in TTMPs. For small values of $A$, the inter-core distance remains large for all values of $|K_\text{eff}|$, whereas for larger values of $A$ the inter-core distance is highly tunable as a function of $|K_\text{eff}|$. Fundamentally, the size scaling for tightly-bound bimerons can depend only on the dimensionless parameter $\kappa$ and the length scale $l_\text{w}$, as introduced in sections \ref{sec:analytics} and \ref{sec:Phenom}. It should be noted that the antimeron component is distorted, causing the bimeron to lose circular symmetry \cite{AsymmSkyrm}.

Whilst our simulations do not address the question of the barrier height directly, we can make some general observations. Assuming a quadratic potential around the equilibrium inter-core (anti)meron separation $s_\text{eq}$, we can approximate the energy of the bimeron (up to quadratic order) as

\begin{equation}
E = \alpha + \beta s + \gamma s^2,
\label{eqn:scaling}
\end{equation}

where $\alpha$, $\beta < 0$ and $\gamma > 0$ are unknown phenomenological parameters that enforce a positive-curvature quadratic with $s>0$. For certain simple ansatz, such as the linear bimeron studied in Appendix \ref{App:BiMeron}, we can identify these three parameters with the micromagnetic parameters $A$, $D_\text{eff}$ and $K_\text{eff}$ respectively, up to some numerical factors. As a result, we can derive the equilibrium separation $s_\text{eq} = -\beta /(2\gamma) \propto D_\text{eff}/K_\text{eff}$ and the barrier height $\Delta = E(0)-E(s) = \beta^2/(4\gamma) = -0.5\beta s_\text{eq} \propto D_\text{eff}^2/K_\text{eff}$. It is clear from our simulation data that the exchange strength $A$ does play a role in determining the bimeron size and therefore likely the barrier height, which is not accounted for in the linear bimeron solution, meaning that the relationship between the phenomenological and micromagnetic parameters is in reality more complicated. Regardless of the exact expression, these considerations indicate that the route towards experimentally realizing closely-bound, stable bimerons is to maximize $\Delta$ and minimize $s_\text{eq}$ at the same time, which requires increasing both $D$ and $K_\text{eff}$. 

\subsection{Comparison with experiments}
When comparing the present results with our recent experiments \cite{Hariom} we are faced with an apparent contradiction. Experimentally, we observed that merons had varied chirality, which seems to rule out the presence of significant iDMI; however, we also observed meron-antimeron pairs that appeared to be quite robust, demonstrating that their lifetimes must be extremely long and implying that the associated energy barrier to annihilation is large. In our simulations, this is only possible in the presence of iDMI. These contrasting observations suggest that an alternative mechanism not accounted for in our micromagnetic model might be responsible in the real system for the apparent stability of these pairs. In terms of the phenomenological model discussed earlier, this means that some additional energy term, other than the exchange, anisotropy and iDMI considered throughout, likely contributes to the barrier height $\Delta$. An alternative explanation is that the potential landscape is locally flat, allowing both bimerons and TTMPs to be trapped by local defects even in the absence of an `intrinsic' potential barrier. This implies that our phenomenological model would need to go beyond the quadratic approximation, such that the inter-core force need not always increase with distance.

For practical implementation of homochiral bimerons in $\alpha \text{-Fe}_2 \text{O}_3$ based racetrack applications, we cannot rely on defects or other local pinning mechanisms to achieve stability because the bimerons must be \emph{mobile}. Therefore, we require bimerons to exist in a local energy minimum at a small inter-core distance $s_\text{eq}$ between the meron and antimeron as well as a large energy barrier $\Delta$ to prevent bimeron annihilation. Our simulations clearly imply that we should be able to engineer the material parameters in such a way as to achieve this goal and that this can only be achieved by topologically protected bimerons (rather than TTMPs) in the presence of a reasonably strong iDMI.

\section{\label{sec:skyrm} Topological skyrmions}

\begin{figure*}
\centering
\begin{subfigure}[t]{0.47\textwidth}
\caption{}
\centering
\includegraphics[width=\textwidth]{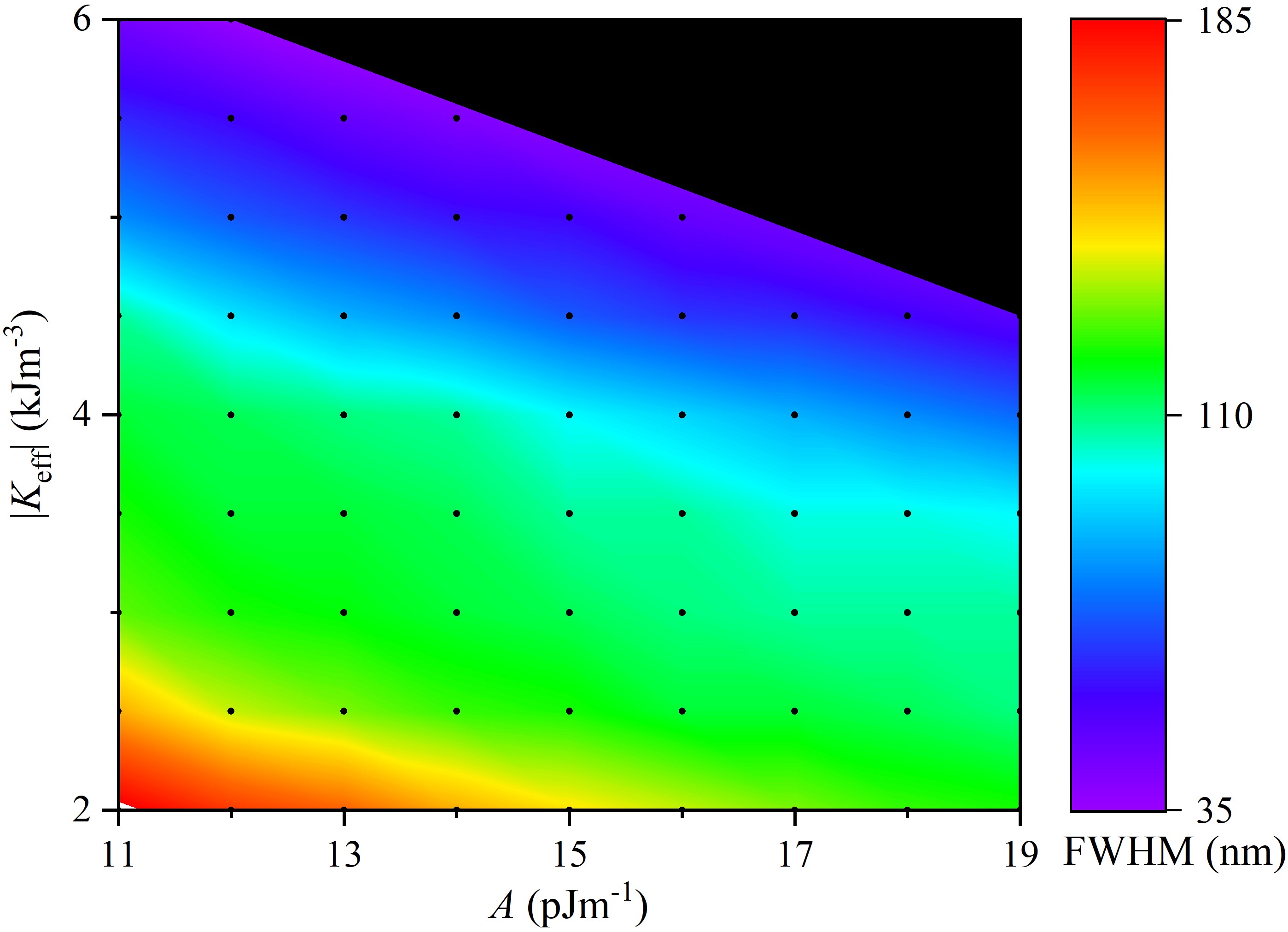}
\label{fig:SkyrmAK}
\end{subfigure}
\hfill
\begin{subfigure}[t]{0.42\textwidth}
\caption{}
\centering
\includegraphics[width=\textwidth]{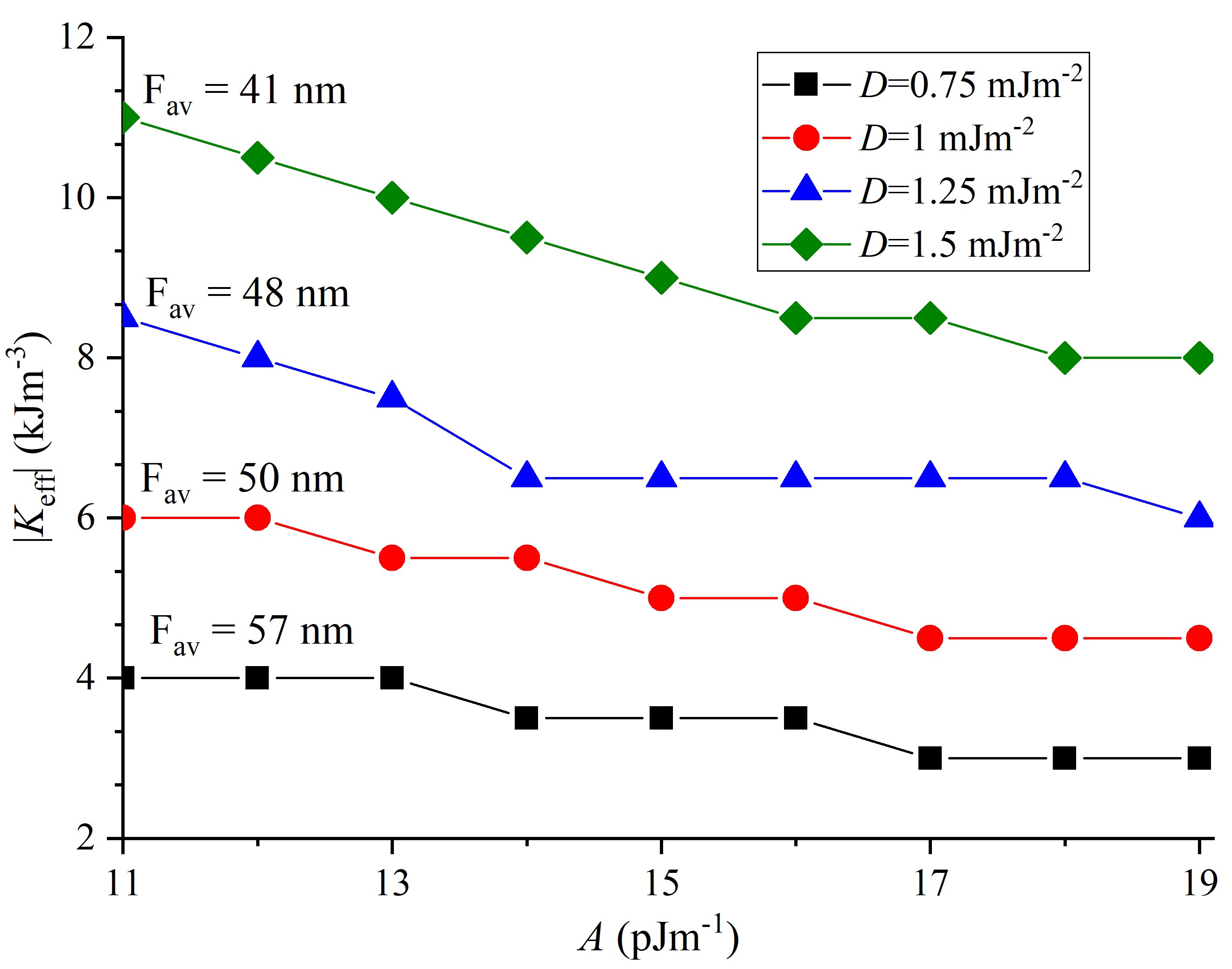}
\label{fig:SkyrmBounds}
\end{subfigure}
\caption{a) N\'{e}el Skyrmion stability window as a function of $A$ and $K_\text{eff}$ for $D$ = 1 mJ\,m$^{-2}$. The color scale represents the relaxed FWHM of the skyrmions and the black area shows the region where the skyrmion was found to radially collapse and therefore was unstable. The $A$ and $K_\text{eff}$ values for which simulations were performed are shown by black squares. b) $K_\text{eff}$ vs $A$ curves at constant skyrmion radius, calculated for the smallest size for which a skyrmion can be stabilized in our simulations (given by the $F_\text{av}$ value next to each line). Symbols/colors correspond to different iDMI strengths.}
\label{fig:Skyrm}
\end{figure*}

Here, we discuss the possibility to stabilize \emph{antiferromagnetic skyrmions} in $\alpha \text{-Fe}_2 \text{O}_3$. These have not been observed experimentally thus far but, as we demonstrate here, are stable in our simulations over a wide range of material parameter values. In $\alpha \text{-Fe}_2 \text{O}_3$, a skyrmion can only exist in the easy-axis phase ($T<T_\text{M}$). It is worth pointing out that the Morin temperature can be raised well above room temperature by chemical doping \cite{Hariom, HDoping}, allowing practical exploitation of such skyrmions. As in the case of bimerons and of skyrmions in other systems without bulk DMI, it is necessary to have a sizable iDMI to stabilise these textures. We therefore initialized a N\'{e}el skyrmion with $K_{\text{eff}}>0$ (i.e. an easy-axis anisotropy along $z$) in the presence of an iDMI. An example of such a skyrmion can be seen in Fig. \ref{fig:Skyrmion} and the full stability window for a range of $A$, $K_{\text{eff}}$ and $D$ values is shown in Fig. \ref{fig:SkyrmAK} and \ref{fig:SkyrmBounds}. For a wide range of micromagnetic parameters, the skyrmion remained stable and either grew or shrunk to an equilibrium size. As expected, these skyrmions are always N\'{e}el type, which is favored by the iDMI. Consequently, when we initialized a Bloch type skyrmion the spins globally rotate into the N\'{e}el configuration. 

We can also study the tuning of physical parameters ($A$ and $K_\text{eff}$) required to minimise the skyrmion size in our system for a given $D$ (see Fig. \ref{fig:SkyrmBounds}). The data in Fig. \ref{fig:Skyrm} points towards a threshold radius, below which the skyrmions spontaneously evaporate via radial collapse \cite{SkyrmCollapse, SkyrmAnni}. This is likely due to a breakdown in the micromagnetic regime (where the finite cell size is on the same order as the length scale of variations). Hence, the sizes shown in Fig. \ref{fig:SkyrmBounds} in fact represent an upper limit to the minimum achievable skyrmion size in this material system, and in fact smaller skyrmions that cannot be reasonably simulated using micromagnetics may also be stable. This shows the potential to generate ultra-small antiferromagnetic skyrmions for practical applications in $\alpha \text{-Fe}_2 \text{O}_3$.

As can be seen in Fig. \ref{fig:SkyrmBounds}, increasing the iDMI strength $D$ or decreasing the exchange coupling $A$ increases the maximum anisotropy $K_\text{eff}$ for which skyrmions are stable. Given the nature of the Morin transition, increasing the strength of $K_{\text{eff}}$ for $T<T_\text{M}$ at fixed $A$ corresponds to reducing the temperature of the system. Therefore, to maximize the thermal stability window for skyrmions in $\alpha \text{-Fe}_2 \text{O}_3$, we need to engineer films with small $A$ and large $D$. Based on our previous data, this is already of the order of 20K for the smallest value of $D$ considered here (see Appendix \ref{App:param}). This is important to understand as applications for AFM skyrmions require a large window of thermal stability. 

To conclude this section, we emphasize a key prediction of our micromagnetic model: the long sought after \emph{antiferromagnetic skyrmion} should be stable and therefore observable in $\alpha \text{-Fe}_2 \text{O}_3$ and potentially other A-type antiferromagnets.

\section{\label{sec:Phenom} Phenomenological scaling}
\begin{figure*}
\centering
\begin{subfigure}[t]{0.47\textwidth}
\caption{}
\centering
\includegraphics[width=\textwidth]{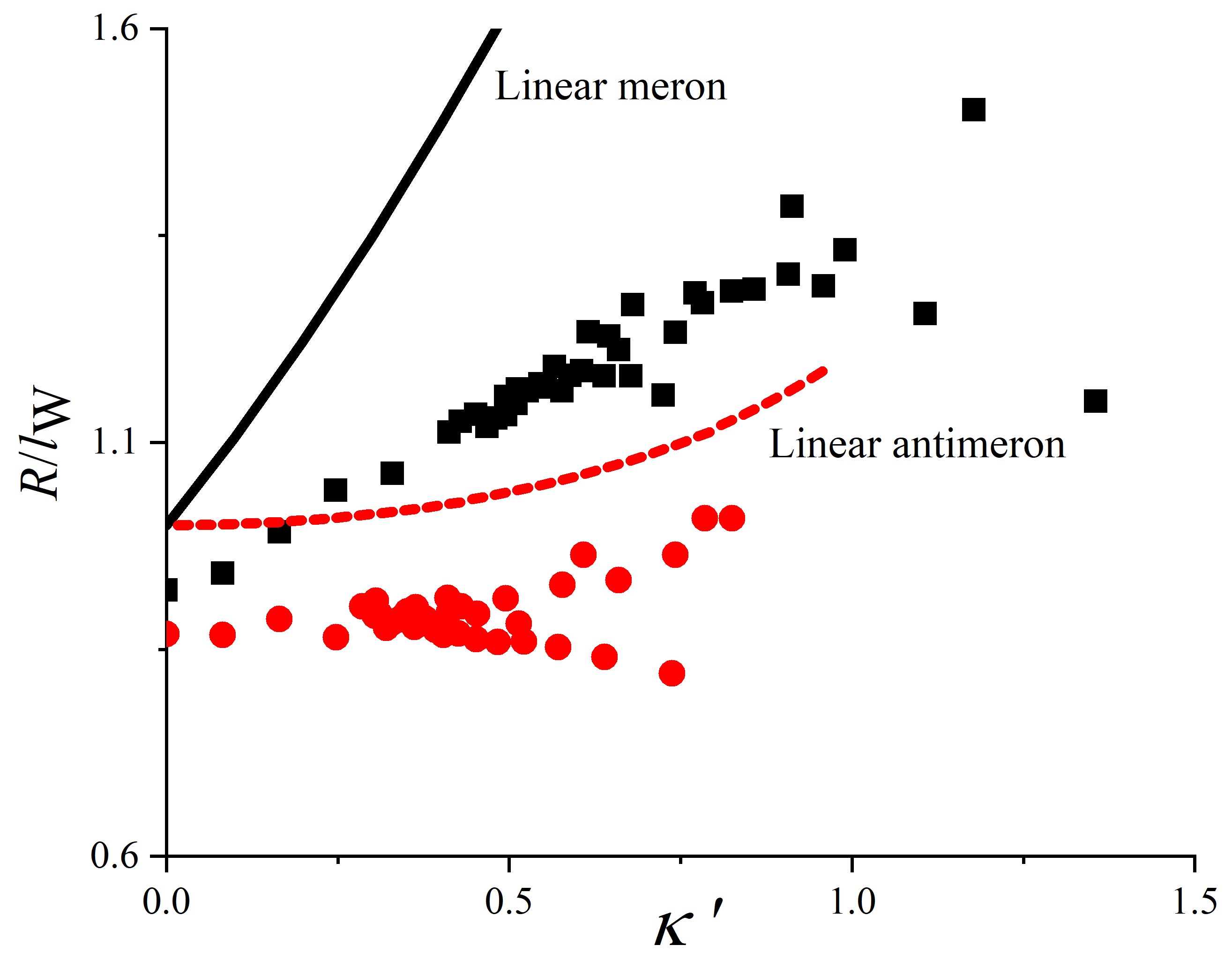}
\label{fig:meronScaling}
\end{subfigure}
\hfill
\begin{subfigure}[t]{0.475\textwidth}
\caption{}
\centering
\includegraphics[width=\textwidth]{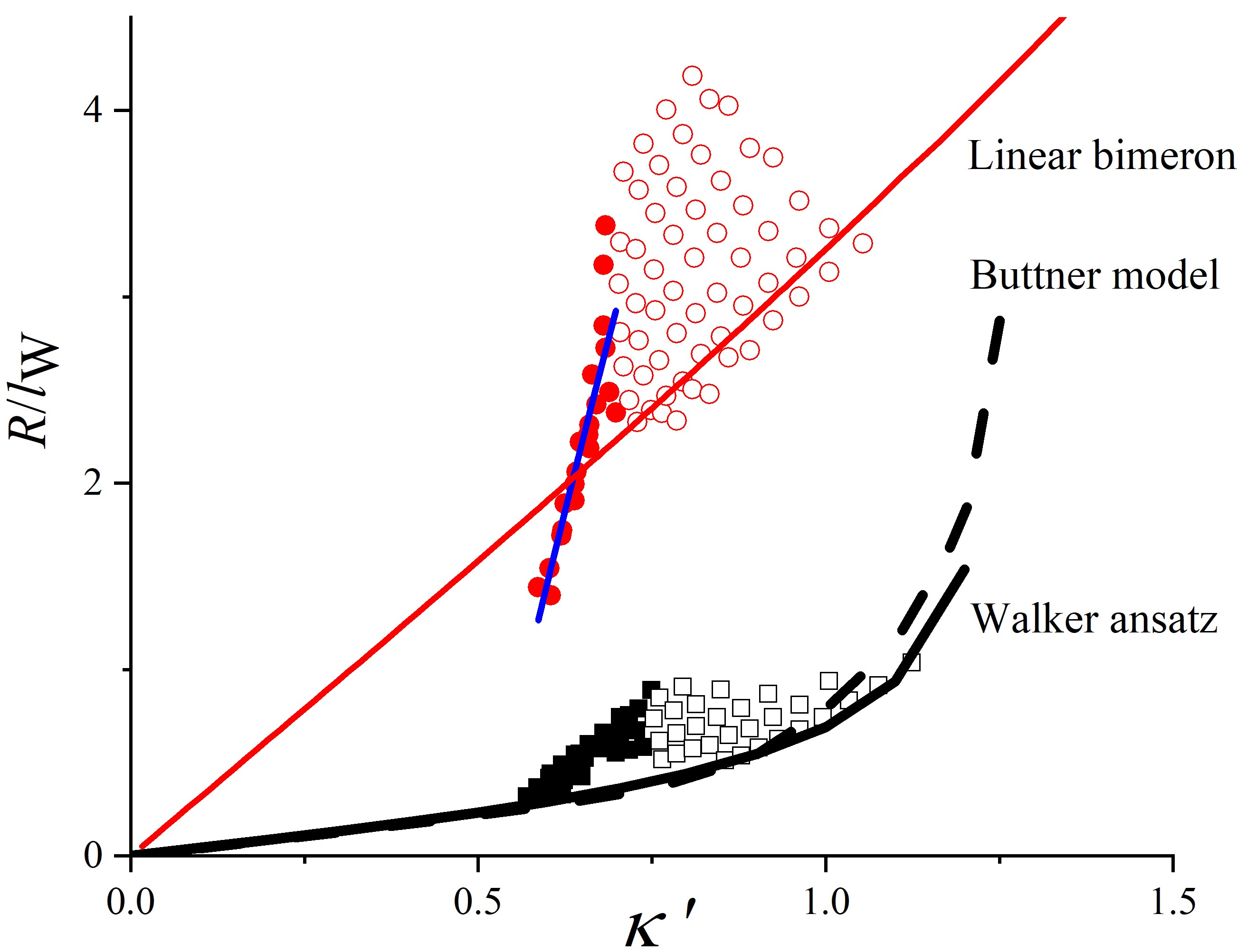}
\label{fig:bimeronScaling}
\end{subfigure}
\caption{a) Rescaled radii of simulated merons (black squares) and antimerons (red circles) as a function of $\kappa'=D_\text{eff}/\sqrt{AK_\text{eff}}$ compared to the corresponding linear (anti)meron ansatz, Eq. \ref{eqn:meron}, \ref{eqn:AMeron1}. b) Rescaled radii of skyrmions (black squares) and bimerons (red circles). In each case the filled-in symbols show the smaller textures in the approximately linear scaling regime and open symbols show larger textures where the energy landscape is very flat. An analytical linear bimeron model and two different skyrmion ansatz \cite{SkyrmModel, SkyrmSize} are shown for comparison. The blue line is a linear fit to the small radii bimerons.}
\label{fig:Scaling}
\end{figure*}

Finally, we discuss the size scaling of various topological textures described herein and demonstrate that our micromagnetic simulations recreate the phenomenological scaling one can expect from simple dimensional analysis. The three materials parameters in our simulations have dimensions $[A] =$ J\,m$^{-1}$, $[K_\text{eff}] =$ J\,m$^{-3}$ and $[D_\text{eff}] =$ J\,m$^{-2}$. We can form a \emph{single} dimensionless parameter $\kappa'=D_\text{eff}/\sqrt{AK_\text{eff}}$ and a length scale $l_\text{w}'=\sqrt{A/K_\text{eff}}$. All length scales in the problem must be proportional to $l_\text{w}'$ multiplied by a dimensionless function of $\kappa'$, this is consistent with our analytical calculations (Eq's \ref{eqn:meron} and \ref{eqn:AMeron1}). Hence, if we were to divide all the relevant sizes of the textures, as extracted from our simulations, by $l_\text{w}'$, we would expect the results to scale as a function of $\kappa'$ only. This is demonstrated in Fig. \ref{fig:Scaling}, where the texture radii are rescaled by $l_\text{w} = \eta \sqrt{A/K_\text{eff}}$, which is the relevant length scale for the linear (anti)meron ansatz as discussed above (see section \ref{sec:analytics}), and then plotted as a function of $\kappa'$. This analysis has two purposes: firstly, to compare the simulated textures to various analytical models presented here and other skyrmion models \cite{SkyrmModel, SkyrmSize} in the literature; secondly, to establish whether or not our simulations have reached equilibrium, which is a precondition for scaling. 

Concerning the first point, for the (anti)meron textures (Fig. \ref{fig:meronScaling}) it is clear that our linear ansatz models reproduce the observed functional scaling to a good approximation, only differing by numerical factors of order one, which is consistent with our discussions throughout. The same is not true for bimerons (Fig. \ref{fig:bimeronScaling}), which clearly display a different functional dependence from the linear approximation (see Appendix \ref{App:BiMeron}), although the sizes are of the correct order of magnitude. We also report the scaling for our AFM skyrmions, which is in good agreement with established skyrmion models in the literature within the range of stability. A straightforward observation is that skyrmions are generally much smaller than bimerons in the whole range of parameters we explored, suggesting that they might be more suitable textures for applications of A-type antiferromagnets. Additionally, whilst many analytical skyrmion models predict a stability threshold for larger $\kappa'$, they fail to predict the lower threshold for $\kappa'$. This is because these are all continuum models that do not account for the atomic nature of the system, which becomes important as the skyrmion approaches smaller length scales.

Turning to the second point, there is a clear distinction between smaller and larger skyrmions/bimerons; whilst smaller textures scale roughly linearly with $\kappa'$, larger textures do not obey any obvious scaling. This is likely because they have not reached their equilibrium radii due to the locally flat energy landscape about the equilibrium (which can be seen by studying the solutions in \cite{SkyrmModel}). In this regime, a range of different sized textures are observed within the simulation tolerances for a given $\kappa'$. We remark that the textures we are interested in for applications, namely those that are small, are also those that fortuitously obey the expected scaling law as a function of $\kappa'$. We also note that this spread of texture sizes for a given $\kappa'$ is not necessarily just an artifact of the simulations, as local strain and thermal fluctuations could also lead to this effect in a real system and this will be most pronounced when the energy landscape is locally flat near the equilibrium.

\section{Summary}
We have presented a comprehensive micromagnetic model for A-type antiferromagnets and applied it to the study of topological textures in $\alpha \text{-Fe}_2 \text{O}_3$ with interfacial DMI. Firstly, we verified our model by comparing simulated N\'{e}el merons and distorted antimerons with corresponding analytical calculations. Given the simplicity of our analytical ansatz, the consistency between the simulated and analytical scaling of these textures ais remarkable. Then, we used our model to analyze the properties of N\'{e}el bimerons in the presence of iDMI and compared their stability and scaling behavior with those in our recent experiments. Finally, we demonstrated that $\alpha \text{-Fe}_2 \text{O}_3$ can also host the long sought-after N\'{e}el \emph{antiferromagnetic skyrmions} and discussed the requirements to experimentally stabilize and observe such textures. We emphasize that our results here demonstrate that a wide family of homochiral topological textures can be stabilized in both the IP and OOP phase of this material, making $\alpha \text{-Fe}_2 \text{O}_3$ an ideal platform for exploring beyond-Moore device architectures exploiting AFM topological textures.

\begin{acknowledgments}
The work done at the University of Oxford was funded by EPSRC grant no. EP/M2020517/1 (Quantum Materials Platform Grant). The work at the National University of Singapore was supported by the National Research Foundation under the Competitive Research Program (NRF2015NRF-CRP001-015).
\end{acknowledgments}

\appendix
\section{\label{App:Eterms} Energy Terms}

For our models we include four energy terms in a continuous form, these being the exchange, uniaxial anisotropy, dipolar and iDMI energies as follows:

\begin{widetext}
\begin{eqnarray}
E_{\text{Ex}} &=& \iiint A \left[ \left(\frac{\partial \bm{\hat{m}}}{\partial x}\right)^2 + \left(\frac{\partial \bm{\hat{m}}}{\partial y}\right)^2 \right] \, d^3r, \label{Ex_int} \\
E_{\text{An}} &=& \iiint -K_\text{eff}\left(\bm{\hat{u}} \cdot \bm{\hat{m}}\right)^2 \, d^3r, \label{An_int} \\
E_{\text{Dip}} &=& \iiint -\frac{1}{2} (\bm{\vec{\mu}} \cdot \bm{\vec{h}}) \,d^3r, \label{Dip_int} \\
E_{\text{DMI}} &=& t \iint D\left(m_x \frac{\partial m_z}{\partial x} - m_z \frac{\partial m_x}{\partial x} + m_y \frac{\partial m_z}{\partial y} - m_z \frac{\partial m_y}{\partial y}\right) \,d^2r. \label{DMI_int}
\end{eqnarray}
\end{widetext}
$A$, $K_\text{eff}$ and $D$ are the exchange, uniaxial anisotropy and iDMI constants respectively. $\bm{\hat{m}}$ is the unit magnetization vector field of the sublattice, $\bm{\hat{u}}$ is the anisotropy axis, $\bm{\vec{\mu}}$ is the local magnetic moment, $\bm{\vec{h}}$ is the local dipolar field and t is the `effective range' of the iDMI (see below) \cite{AmikamBook, iDMIEnergy}. The micromagnetic simulations utilize a discretized version of these equations \cite{mumax1, mumax2, mumax3}. In our model, an exchange interaction of the form in Eq. \ref{Ex_int} only applies to the ferromagnetic interactions between adjacent cells in the same layer (we recall that the inter-layer exchange strictly enforces antiparallel alignment between adjacent layers in this model). Note that $K_\text{eff}>0$ yields an easy-axis parallel to $\bm{\hat{u}}$ whereas $K_\text{eff}<0$ yields an easy-plane perpendicular to $\bm{\hat{u}}$. Eq. \ref{DMI_int} requires the iDMI energy to be uniform throughout a thickness $t$ \cite{iDMIEnergy}, which we assume to correspond to the thickness of a single AFM layer. This is different to the other three energy terms, which are volume integrals, whereas the iDMI contribution comes from a surface integral. This implies no loss of generality, provided we assume that the textures are not modulated along $\bm{\hat{z}}$.

\section{ \label{App:param} Micromagnetic parameters of $\alpha \text{-Fe}_2 \text{O}_3$}

In our simulations, the cell size along the $z$-axis was fixed at 0.228 nm, corresponding to $1/6$ of the unit cell and the spacing between AFM layers \cite{CoeyBook}. The simulation size and cell size was adjusted for each different texture type to ensure a good compromise between simulation time and texture scales in each case, given that the skyrmion and bimeron simulations took an order of magnitude longer than the (anti)merons. In all cases, different cell dimensions and total simulation dimensions were checked and found to be consistent and all satisfied the micromagnetic guideline of a maximum angle between adjacent moments of no more than $20^{\circ}$ (in the x-y plane), hence the choices made throughout are purely a matter of convenience.

The sublattice saturation magnetization of $\alpha \text{-Fe}_2 \text{O}_3$ is 920 kA\,m$^{-1}$ \cite{CoeyBook}. The IP exchange constant $A$ in $\alpha \text{-Fe}_2 \text{O}_3$ (i.e. the ferromagnetic interaction between cells in the same layer) is around 14-17 pJ\,m$^{-1}$\cite{Hariom} depending on the exchange parameters used to calculate it \cite{CoeyBook, HematiteJvals} and this can be altered further via doping or strain \cite{RhDoping, FeStrain}, justifying the range of values $A$ = 10-20 pJ\,m$^{-1}$ used here. The AFM coupling between adjacent layers was similarly calculated to be $A_{\text{OOP}} = -20.1$ pJ\,m$^{-1}$ and is kept constant throughout. As the long-range dipolar fields are negligible, the corresponding $z$-axis magnetic exchange length $l_{ex} = \sqrt{2A/(\mu_0m_s^2)}$ (where, $m_s$ is the weak canted ferromagnetic moment) will be much larger than the simulation size along $z$, so we expect negligible texture variation in this dimension (consistent with all our simulations). We have additionally confirmed that altering the cell size along $z$ has no discernible effect on the textures, thereby further justifying our approach. 

In $\alpha \text{-Fe}_2 \text{O}_3$, the effective anisotropy constant $K_\text{eff}$ results from a competition of on-site and dipolar interactions \cite{MorrishBook, BPAnis, FeAnis}, with the Morin transition $T_\text{M}$ occurring when these two interactions balance. This competition can be tuned by strain, chemical doping and reversible ionic control to alter $T_\text{M}$ or destroy the transition altogether \cite{FeStrain, BPAnis, HDoping}. As both the on-site and dipolar anisotropies are temperature-dependent, the value of $K_\text{eff}$ varies systematically either side of $T_\text{M}$ and these values can be calculated and directly compared with our simulation data. For example, using the representative values $A = 14$ pJ\,m$^{-1}$ and $D = 0.75$ mJ\,m$^{-2}$, the maximum anisotropy value for which we observed stable skyrmions was $K_\text{eff} = 3.5$ kJ\,m$^{-3}$ (see Fig. \ref{fig:SkyrmBounds}). Based on a thin film with $T_\text{M} =$ 240 K, similar to that used in our previous experiments \cite{Hariom}, this corresponds to a temperature of approximately 219 K, meaning that we estimate the skyrmion stability window to be on the order of 20 K below $T_\text{M}$, which is certainly a feasible range for practical observation of these textures. As we are not aware of any work studying or engineering possible iDMI strengths in $\alpha \text{-Fe}_2 \text{O}_3$ systems, we have used values of $D$ throughout that are akin to those used in other theoretical studies \cite{AFMSkyrm1, iDMIanatomy, SkyrmModel} and to those found in the large body of work on topological texture hosting Co-Pt heterostructures (\cite{CoDMI}).

As briefly outlined in section \ref{sec:General}, the dipolar fields can be approximately decomposed into two parts: the short-range dipole interaction, which is generally taken to contribute to the anisotropy energy and is therefore not included in the model separately, and the long-range part, which is computed directly in the form of a demagnetization field \cite{AmikamBook}. In a natural collinear antiferromagnet, the net magnetization is zero and so is the demagnetization field, but the short-range component of the dipolar interaction still contributes to the anisotropy. In our `model' A-type antiferromagnet, each magnetic cell has an exactly antiparallel counterpart in a vertically adjacent layer. Therefore, the calculated demagnetization field has a quadrupolar character and decays very rapidly. Although there are no macroscopic demagnetization fields and no shape anisotropy, the short-range part of the quadrupolar interaction still generates an anisotropy, inducing a strong preference for the moments to lie in the $x$-$y$ plane even when the on-site anisotropy energy density is set to zero. This situation is very similar to that of `real' $\alpha \text{-Fe}_2 \text{O}_3$, since the Morin transition occurs precisely when the easy-axis on-site anisotropy exactly balances the perpendicular easy-plane dipolar anisotropy. As our simulations use a cuboidal configuration rather than stacked honeycomb layers, we would not expect the dipolar anisotropy to be exactly the same as for the real material, but the order of magnitude should be correct. In fact, our calculated dipolar anisotropy is approximately 60\% of the known value for $\alpha \text{-Fe}_2 \text{O}_3$. For now, we observe that our `model' A-type antiferromagnet provides a good physical account of the real material, provided that the dipolar anisotropy calculated by the micromagnetic code is properly taken into account.

\section{\label{App:Analytics} Analytical Calculations}

In polar coordinates, the linear (anti)meron ansatz for the unit magnetization vector $\bm{\hat{m}}=(\sin \theta \cos (\phi + \xi), \pm \sin \theta \sin (\phi + \xi), \cos \theta)$ is

\begin{equation}\label{LMAnsatz}
\theta (r) = \begin{cases}
	{\pi r}\over{2R} & \text{for r $\leq$ R} \\
	{\pi}\over{2} & \text{for r $>$ R}, \\
\end{cases}
\end{equation}

where the $+$ ($-$) sign corresponds to a meron (antimeron) and $R$ is the \lq{}(anti)meron radius\rq{}, representing the typical size of our texture. The angle $\phi$ is the in-plane azimuthal angle whereas $\xi$ is an additional phase angle that determines the overall chirality. The dipolar interaction will only be considered as a contribution to the anisotropy, so energy terms of the form in Eq. \ref{Dip_int} will not appear explicitly in the analytical calculations below. As all the calculations herein will be for the above-Morin state with an easy-plane anisotropy ($K_{\text{eff}}<0$), we will drop the `$-$' sign in Eq. \ref{An_int} above and the absolute value $|K_{\text{eff}}|$ will be used throughout and designated as $K$ for simplicity.

\subsection{\label{App:Meron} Analytical Meron}

We start by studying a linear meron, which is an extension of a calculation we have done previously \cite{Jacopo, Hariom}, but with an iDMI term now included. We showed therein that the exchange and anisotropy energy terms are independent of any phase angle $\xi$. This is not the case for the iDMI energy, which contains a term that explicitly depends on $\xi$. Using Eq. \ref{DMI_int} the resulting integral is

\begin{equation}
E_{\text{DMI}} = -Dt\cos \xi\iint\left(\frac{\pi r}{2R} + \cos \theta \sin \theta \right) \, drd\phi.
\end{equation}

As the first term comes from $\partial \theta/\partial r$, it is only present for $r \leq R$. Moreover, it can be easily seen that for $r > R$, $\theta=\pi/2$ and hence $\cos \theta = 0$, so we only need to integrate the above expression in the range $0 < r \leq R$. Due to the axial symmetry, the integrand is independent of $\phi$, hence the angular integral results in a factor of 2$\pi$. The radial integral gives

\begin{equation}\label{iDMI_Ef}
E_{\text{DMI}} = -\frac{1}{2}DRt\cos \xi \left(\pi^2+4\right).
\end{equation}

The exchange and anisotropy energies from our previous calculations \cite{Hariom} are:

\begin{eqnarray}
E_{\text{Ex}} &=& 2\pi ANt\left[C-\ln\left(\frac{R}{a}\right)\right] \label{Ex_Ef}, \\
E_{\text{An}} &=& \frac{\pi^2-4}{2\pi} KNtR^2 \label{An_Ef},
\end{eqnarray}

where $C \approx 2$ is a numerical constant. $N$ is the number of AFM layers of thickness $t$, such that $Nt$ is the total thickness of the film and $a$ is a small limit introduced to remove the infinite exchange energy contribution from the extended whirling IP background by subtracting off a flat vortex and will later be set to zero (see \cite{Jacopo}). We can then combine Eq's. \ref{iDMI_Ef}, \ref{Ex_Ef} and \ref{An_Ef} to get the total energy of an analytical linear meron

\begin{widetext}
\begin{equation}
E_T = 2\pi ANt\left[C-\ln\left(\frac{R}{a}\right)\right]+\frac{\pi^2-4}{2\pi} KNtR^2 - \frac{1}{2}DRt\cos \xi \left(\pi^2+4\right).
\end{equation}
\end{widetext}

As only the iDMI energy term has a chiral component (i.e. depends on $\xi$), if we minimize \ref{iDMI_Ef} with respect to $\xi$ we can find the equilibrium chirality; $\xi = 2n\pi$ for integer $n$ and positive $D$ or $\xi=(2n+1)\pi$ for negative $D$. We take $\xi=0$ for convenience, which corresponds to a N\'{e}el meron of a fixed chirality and is an energy minimum under the assumption of a positive D, as is used in the simulations. We rescale the equation by the film thickness and define an effective DMI strength $D_{\text{eff}} = D/N$. Minimizing the total energy with respect to $R$ we can find the equilibrium meron radius

\begin{widetext}
\begin{equation}
R = \frac{\pi}{2K(\pi^2-4)}\left[\frac{1}{2}D_{\text{eff}}(\pi^2+4) \pm \sqrt{\frac{1}{4}D_{\text{eff}}^2(\pi^2+4)^2 + 8AK(\pi^2-4)}\right].
\end{equation}
\end{widetext}

In the limit of no iDMI or of thick films (such that $D_{\text{eff}}\rightarrow 0$), the result above reduces to that found previously in the absence of iDMI \cite{Hariom}. Furthermore, as all the terms in the square root are positive with our conventions, the `$-$' sign in the above expression would give a negative radius, which is unphysical and is therefore eliminated. As the $z$-component of the normalized magnetization is given by $\cos(\pi r/2R)$, we find that $m_z = 0.5$ when $r = 2R/3$, hence comparing the meron radius $R$ to the FWHM $F$ we obtain $F = \frac{4}{3}R$. 

To compare our results to previous analytical studies, e.g. \cite{SkyrmModel}, we introduce the characteristic length scale $l_w$ and a dimensionless parameter $\kappa$ defined as

\begin{eqnarray}
l_w &=& \pi \sqrt{\frac{2A}{(\pi^2-4)K}}= \eta \sqrt{\frac{A}{K}}, \\ \nonumber
\kappa &=& \frac{(\pi^2+4)}{\left[4\sqrt{2(\pi^2-4)}\right]}*\frac{D_{eff}}{\sqrt{AK}} = \kappa_0 \frac{D_{eff}}{\sqrt{AK}}.
\label{eqn:params}
\end{eqnarray}

$l_w$ is equivalent to the meron radius when $D_{\text{eff}} \rightarrow 0$ and $\kappa$ is the unique dimensionless parameter that can be formed given the parameters involved, up to numerical factors (see section \ref{sec:Phenom}). This allows us to express the meron radius in a simplified form

\begin{equation}
R = l_w \left(\kappa + \sqrt{\kappa^2 + 1}\right).
\end{equation}

\subsection{\label{App:AntiMeron} Distorted Antimeron}

As an antimeron is composed of both N\'{e}el and Bloch sectors, we expect that it should distort in the presence of iDMI. We approach this analytically in a similar way to the meron studied above and use a linear antimeron ansatz where $\bm{\hat{m}}=(\sin \theta \cos \phi, -\sin \theta \sin \phi, \cos \theta)$ and $\theta(r)$ has the same functional form as before (Eq. \ref{LMAnsatz}). We note that the sign of $\bm{\hat{m}}_y(r,\phi)$ is reversed compared to the meron and observe that here it is not necessary to include a phase factor $\xi$, as this would only lead to a global rotation of the antimeron. This can be easily shown by considering a 3D rotation matrix about the $z$-axis ($R_z$) by an angle $\xi$ applied to $\bm{\hat{m}}$, i.e. $R_z(\xi)\bm{\hat{m}}$. Moreover, we treat the distortions by introducing two additional parameters, $\lambda$ and $\mu$, which modify the mapping between $(x, y)$ and $(r, \phi)$ so that constant-$r$ lines are ellipses rather than circles. We write $r = \sqrt{(\lambda x)^2 + (\mu y)^2}$ and $\tan \phi = \mu y/(\lambda x)$, such that $\lambda x = r \cos \phi$ and $\mu y = r \sin \phi$. We note at this stage that such a distortion should depend on only a single parameter; as a result, we will later enforce the criteria $\lambda\mu = 1$, which corresponds to requiring that all equal-$r$ ellipses have the same area. We then use Eq's \ref{Ex_int}, \ref{An_int} and \ref{DMI_int} to calculate how the relevant energy terms are modified by these distortions. Firstly, for textures that do not vary along the z-direction the exchange energy contribution is

\begin{equation}\label{Ex_int2}
E_{\text{Ex}} = \iint ANt \left[ \left(\frac{\partial \hat{m}}{\partial x}\right)^2 + \left(\frac{\partial \hat{m}}{\partial y}\right)^2 \right] \, d^2r.
\end{equation}

Most of this calculation proceeds in the same way as for a meron, requiring only slight modifications to introduce factors of $\mu$ and $\lambda$. The resulting exchange energy integral is

\begin{widetext}
\begin{equation}
E_{\text{Ex}} = ANt \iint \left[ \left(\frac{\pi}{2R}\right)^2\left(\lambda^2\cos^2 \phi +\mu^2 \sin^2 \phi \right) +\frac{1}{r^2}\sin^2 \theta \left(\lambda^2 \sin^2 \phi +\mu^2 \cos^2 \phi \right) \right] \, d^2r.
\end{equation}
\end{widetext}

As the first term in the above integral comes from $\partial \theta/\partial r$ it is only non-zero in the range $0 \leq r \leq R$ and is zero outside of this range. The second term is in principle non-zero for all $r$, but needs to be split into two cases corresponding to the two ranges of $r$. Most of these integrals are simple, the radial integral for the second term gives

\begin{equation}
\int_0^{R_\text{d}} \frac{1}{r} \sin^2 \theta dr = C\rq{} + ln \left(\frac{R_\text{d}}{R}\right),
\end{equation}

where $C\rq{}$ is a numerical constant. $R_\text{d}$ is an upper limit introduced to avoid the infinite energy contribution from the whirling background as $r \rightarrow \infty$ that is an artifact of the analytical approach. This effectively corresponds to a long-scale relaxation of the spins away from the whirling background due to the antimeron at a radius $R_\text{d}$, which should have no real effect on the texture itself provided $R_\text{d} \gg R$. This is different to the approach we have taken previously \cite{Hariom, Jacopo} (where we subtracted the energy of an infinite flat vortex to remove the infinite exchange energy), but is more easily interpreted in the framework of micromagnetics and gives the same final result. Hence, the exchange energy of the distorted antimeron is

\begin{equation}\label{Ex_Ef2}
E_{\text{Ex}} = \pi ANt \left(\frac{\lambda^2 + \mu^2}{\mu \lambda}\right) \left[ C+ ln \left(\frac{R_\text{d}}{R}\right) \right],
\end{equation}

where the numerical constant $C \approx 2$ is identical to the constant appearing in the exchange energy of a meron (see Eq. \ref{Ex_Ef}). Next we calculate the anisotropy energy, which has the general form given in \ref{An_int} with $\bm{\hat{u}}=\bm{\hat{z}}$, i.e.

\begin{equation}
E_{\text{An}}=KNt\int_0^{2\pi} \int_0^{R_\text{d}} \cos^2 \theta \, \frac{r}{\mu \lambda} dr d\phi.
\end{equation}

We again introduce an upper limit $R_\text{d}$ and consider the two relevant regimes. For $r > R$, $\theta = \pi/2$ and $\cos \theta = 0$, hence we only integrate up to $R$ and the anisotropy energy is independent of the cutoff radius (provided $R_\text{d} \gg R$). The resulting anisotropy energy is

\begin{equation}\label{An_Ef2}
E_{\text{An}} = KNtR^2 \left(\frac{\pi^2-4}{2\pi \mu \lambda}\right).
\end{equation}

Finally we calculate the iDMI energy using \ref{DMI_int} and the method in section \ref{App:Meron}. The resulting expression is

\begin{widetext}
\begin{equation}
E_{\text{DMI}} = Dt \iint \left[\frac{\pi}{2R}\left(\mu \sin^2 \phi - \lambda \cos^2 \phi \right) + \frac{1}{r}\sin \theta \cos \theta \left(\mu \cos^2 \phi - \lambda \sin^2 \phi \right) \right] \, \frac{r}{\mu \lambda} drd\phi.
\end{equation}
\end{widetext}

As before, the terms in this expression will be zero for $r > R$, hence we only integrate up to $r = R$. We note that the iDMI energy is also independent of $R_\text{d}$ and is calculated to be

\begin{equation}\label{iDMI_Ef2}
E_{\text{DMI}} = DtR\left(\frac{\mu - \lambda}{\mu \lambda}\right)\left(\frac{\pi^2}{4}+1\right).
\end{equation}

By combining \ref{Ex_Ef2}, \ref{An_Ef2} and \ref{iDMI_Ef2} we get the total energy of a distorted antimeron

\begin{widetext}
\begin{equation}
E_T = Nt\left\{\pi A \left(\lambda^2+\frac{1}{\lambda^2}\right) \left[ C+ ln \left(\frac{R_\text{d}}{R}\right) \right] + KR^2 \left(\frac{\pi^2-4}{2\pi}\right) - \frac{1}{4}D_{\text{eff}}R \left(\lambda - \frac{1}{\lambda}\right) \left(\pi^2+4\right)\right\},
\end{equation}
\end{widetext}

where we have enforced the criteria $\lambda\mu = 1$ as discussed previously and thereby expressed the distortions purely in terms of $\lambda$. We note that all three energy terms scale with a factor $t$, the thickness of each AFM layer, hence the equilibrium properties of the antimeron will be independent of this parameter. We now want to find the equilibrium properties of the antimeron by taking the derivatives of the above expression for the total energy with respect to $R$ and $\lambda$. We can then rearrange these expressions and solve the quartic equation resulting from them to get the equilibrium radius and distortion respectively:

\begin{widetext}
\begin{eqnarray}
R &=& \frac{\pi}{2K(\pi^2-4)}\left[\frac{1}{4}D_{\text{eff}}\left(\lambda - \frac{1}{\lambda}\right)(\pi^2+4) + \sqrt{\frac{1}{16}D_{\text{eff}}^2\left(\lambda - \frac{1}{\lambda}\right)^2(\pi^2+4)^2 +4AK\left(\lambda^2+\frac{1}{\lambda^2}\right)(\pi^2-4)}\right], \\
\lambda &=& \frac{D_{\text{eff}}R(\pi^2+4)}{16\pi A\left[ C+ ln \left(\frac{R_\text{d}}{R}\right) \right]} + \sqrt{1+\left\{\frac{D_{\text{eff}}R(\pi^2+4)}{16\pi A\left[ C+ ln \left(\frac{R_\text{d}}{R}\right) \right]}\right\}^2}.
\end{eqnarray}
\end{widetext}

These coupled equations do have an exact solution for $R$ and $\lambda$, however it is easier to calculate these iteratively given values for the other parameters. If $D_{\text{eff}} \rightarrow 0$, the antimeron radius reduces to that found in the case of an (anti)meron without iDMI \cite{Hariom}. As discussed earlier, the cutoff radius $R_\text{d} \gg R$ is an artifact of the analytical approach reflecting the fact that the elliptical antimeron effectively extends to $\infty$, thereby providing a divergent energy contribution. In reality, the spins far from the antimeron would relax into some other configuration that contributes a finite energy and, as no system is truly infinite, this divergent energy is not an issue either practically or in our simulations. For all relevant parameter ranges it would be reasonable to set the cutoff radius as proportional to the exchange length and on these scales varying $R_\text{d}$ by an order of magnitude has little effect on the actual antimeron distortion when calculated using the iterative scheme. If we re-express these formulae in terms of the same dimensionless parameter $\kappa$ and length scale $l_w$ as for the meron (Eq. \ref{eqn:params}), the expressions simplify slightly to

\begin{widetext}
\begin{eqnarray}
R &=& l_w\left[\frac{1}{2}\kappa \left(\lambda - \frac{1}{\lambda}\right) + \sqrt{\frac{1}{4}\kappa^2 \left(\lambda - \frac{1}{\lambda}\right)^2 + \frac{1}{2}\left(\lambda^2+\frac{1}{\lambda^2}\right)}\right], \\
\lambda &=& \frac{\kappa R}{2l_w\left[C+ ln \left(\frac{R_\text{d}}{R}\right) \right]} + \sqrt{1 + \left\{\frac{\kappa R}{2l_w\left[C+ ln \left(\frac{R_\text{d}}{R}\right) \right]}\right\}^2}.
\end{eqnarray}
\end{widetext}

\subsection{\label{App:BiMeron} Distorted linear bimeron}

Finally, we study a linear bimeron ansatz, which can be viewed as a linear skyrmion rotated by $90\degree$ about any IP axis. The form of a linear skyrmion is similar to the linear meron studied earlier, except Eq. \ref{LMAnsatz} is replaced with

\begin{equation}\label{LSkyrm}
\theta (r) = \begin{cases}
	{\pi r}\over{R} & \text{for r $\leq$ R} \\
	\pi & \text{for r $>$ R}, \\
\end{cases}
\end{equation}

which ensures that the magnetization is OOP at $r=0$ and for $r > R$. The magnetization of a linear bimeron is then given by $\bm{\hat{m}}=(-\cos \theta, \sin \theta \sin (\phi + \xi), \sin \theta \cos (\phi + \xi))$. For generality, we allow distortions of the form $r = \sqrt{(\lambda x)^2 + (\mu y)^2}$ as we did for the antimerons discussed above. We can then calculate the relevant energy terms, summarized below:

\begin{eqnarray}\label{BimEnergies}
E_\text{Ex} &=& C''\pi ANt \left(\frac{\lambda^2+\mu^2}{\lambda \mu}\right), \\
E_\text{An} &=& \frac{\pi KNt R^2}{4\lambda \mu}, \\
E_{\text{DMI}} &=& \frac{-\pi^2}{2\mu} DtR,
\end{eqnarray}

where all the symbols have the same meaning as before and $C'' \approx 6$ is a numerical constant. Note that, unlike for an (anti)meron, the exchange energy is independent of the radius $R$ as the background is uniform and there is no extended whirling structure. As before, we can minimize the total energy as a function of $\lambda$ and $R$ after enforcing the constraint $\lambda \mu = 1$ to get the equilibrium distortion and radius satisfying the equations

\begin{eqnarray}\label{BimSolution}
R &=& \frac{\pi \lambda D_\text{eff}}{K} \propto l_w \kappa \lambda\\
0 &=& AC'' \left(2\lambda - \frac{2}{\lambda^3}\right) -\frac{1}{2}\pi D_\text{eff} R.
\end{eqnarray}

Like the antimeron, these equations do have an exact solution for any given set of parameters, however the numerical solutions are more useful, plotted in Fig. \ref{fig:bimeronScaling}. This solution clearly does not reproduce the simulation data overly well, especially at low $\kappa'=D_\text{eff}/\sqrt{AK_\text{eff}}$; the simulated bimerons are not stable in this regime whereas they are stable down to $\kappa'=0$ under the linear ansatz.

\section{\label{App:NoDMI} Comparing simulated merons without iDMI to analytics and data.}

\begin{figure}
\centering
\includegraphics[width=0.45\textwidth]{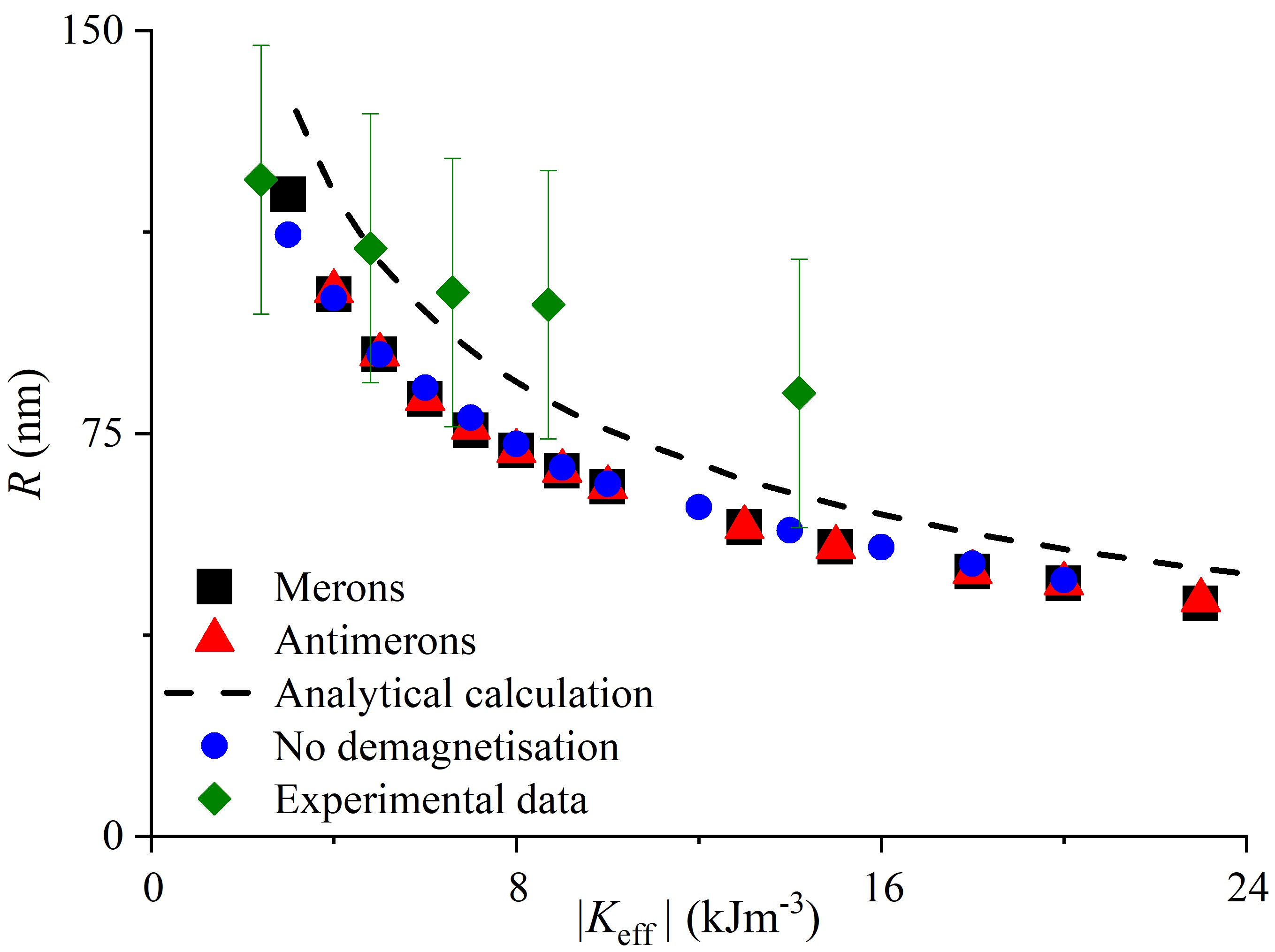}
\caption{The scaling of a meron (black squares) and an antimeron (red triangles) as a function of the effective anisotropy in the absence of iDMI. The dashed line is an analytical calculation using Eq. \ref{eqn:meron} in the limit $D_{\text{eff}} \rightarrow 0$. The blue circles show the radius of simulated merons without iDMI and with the demagnetizing fields turned off, so that the applied easy-plane anisotropy is of strength $K_\text{os}$. The green diamonds show experimental data from \cite{Hariom}, with functional scaling similar to that found in both our simulations and calculations.}
\label{fig:MeronNoDMI}
\end{figure}

\begin{figure*}
\centering
\begin{subfigure}[t]{0.32\textwidth}
\centering
\includegraphics[width=\textwidth]{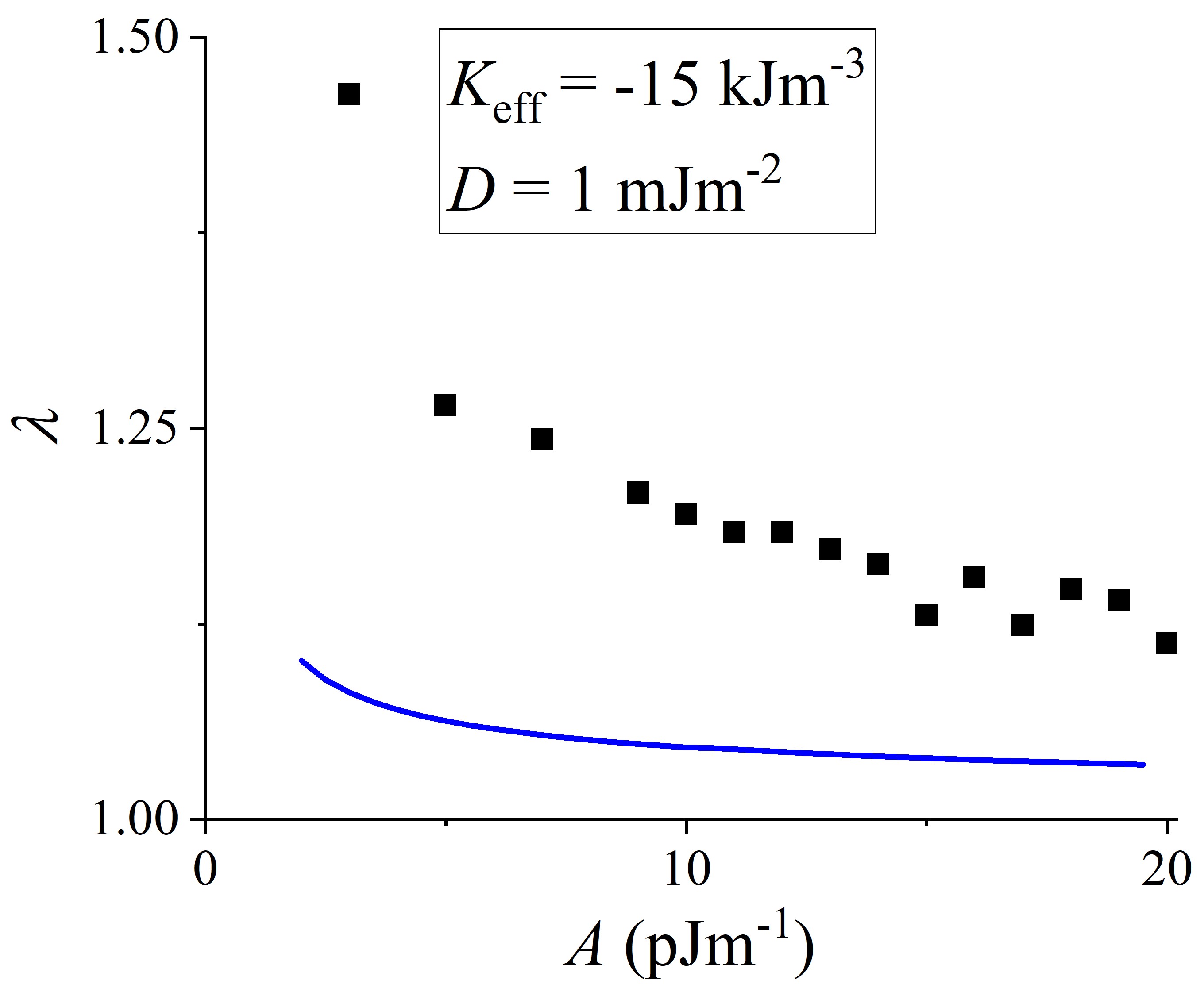}
\caption{}
\end{subfigure}
\hfill
\begin{subfigure}[t]{0.32\textwidth}
\centering
\includegraphics[width=\textwidth]{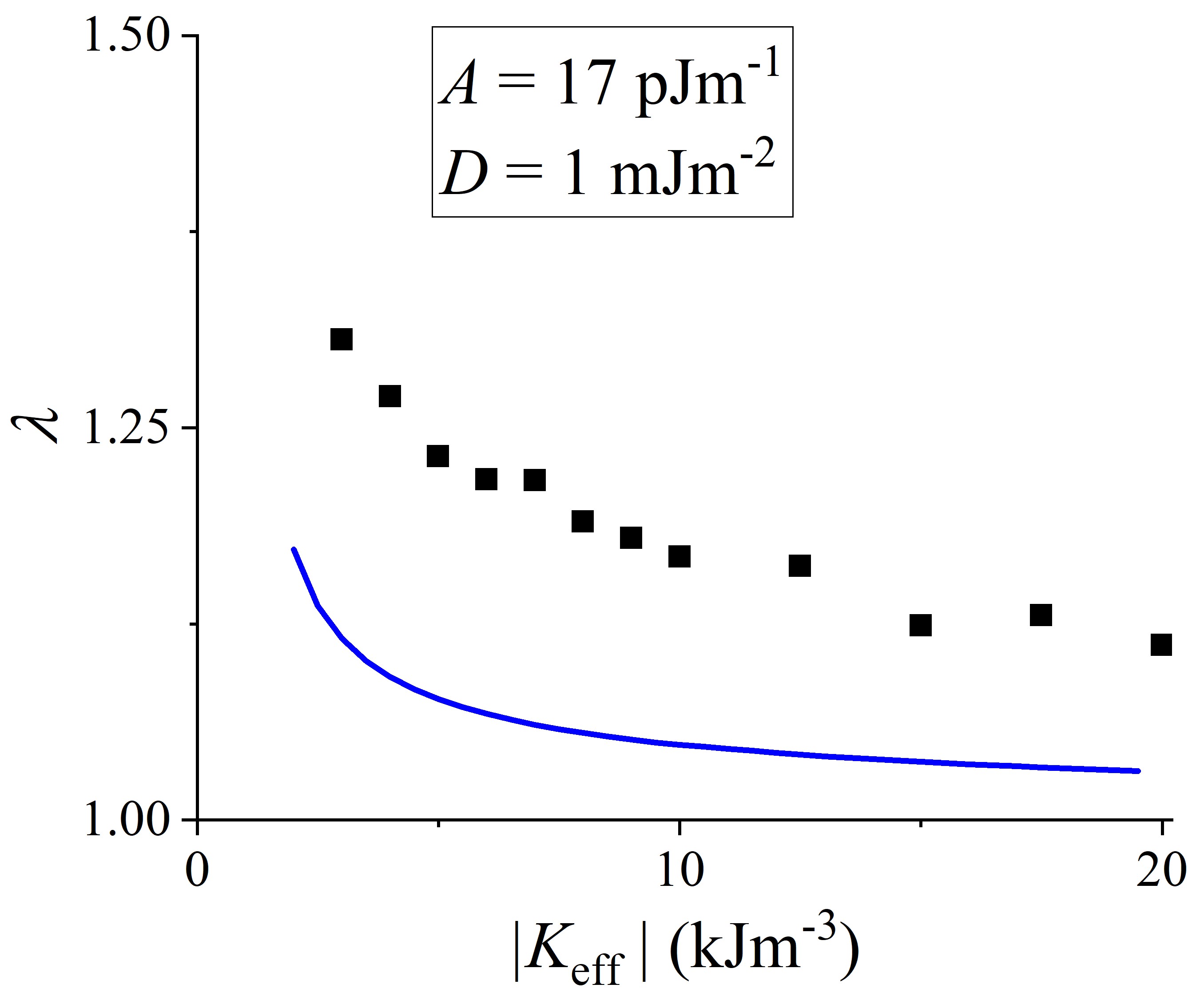}
\caption{}
\end{subfigure}
\hfill
\begin{subfigure}[t]{0.32\textwidth}
\centering
\includegraphics[width=\textwidth]{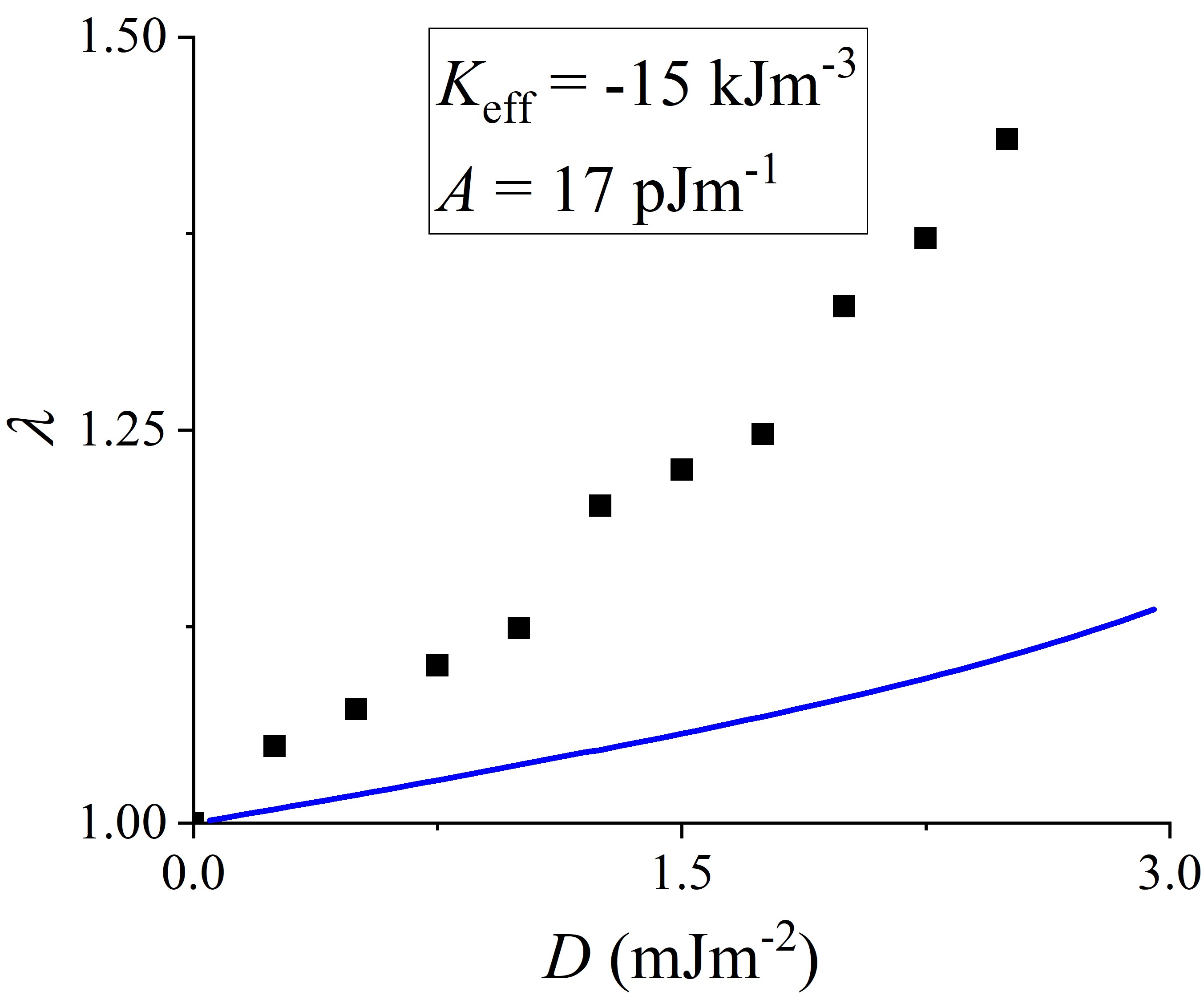}
\caption{}
\end{subfigure}
\caption{Antimeron distortion parameter ($\lambda$) based on micromagnetic simulations (black points) and analytical calculations (blue lines). $A$, $K$ and $D$ were varied in panels (a), (b) and (c) respectively, with the rest of the parameters kept constant.}
\label{fig:AMDist}
\end{figure*}

To validate our micromagnetic model, we performed a set of simulations for merons and antimerons without iDMI that were compared against experimental data \cite{Hariom} and analytical calculations, see Fig. \ref{fig:MeronNoDMI}. It's evident that in the absence of iDMI the radii of merons and antimerons are effectively identical, this is also clear from the analytical expressions discussed in Appendix \ref{App:Analytics}. By considering Eq. \ref{eqn:meron} and taking the case $D=0$ we can also compare this data to the analytical fit, where we see that the scaling matches well but the exact size is off by a similar amount to the cases with iDMI.

We also studied (anti)merons without the demagnetizing field so as to assess whether there are any large effects due to the dipolar fields other than the easy-plane anisotropy contribution as discussed in Appendix \ref{App:param}. Here, the anisotropy is provided by a negative coefficient on-site anisotropy constant $-K_\text{os}$ along the $z$-axis of the same effective strength as for the situation with dipolar fields, where the total anisotropy is given by $K_{\text{eff}} = K_\text{os} - K_{\text{dip}}$. It is clear that these two cases are equivalent, so we are confident in our assessment that incorporating the dipolar fields organically in our model effectively results in an easy-plane anisotropy of the correct order of magnitude and that there are no other major effects of such fields on our magnetic textures.

\section{\label{App:DistGraphs} Comparing antimeron distortions in simulations and analytics}

Following the discussion in section \ref{sec:AMeron}, we present here the comparison between the distortion of a simulated antimeron and that expected from the analytical calculations, as expressed in Eq. \ref{eqn:AMeron2}. These can be seen in Fig. \ref{fig:AMDist} for the same set of antimeron simulations as presented in Fig. \ref{fig:AntiMeron}. Overall, it is clear that the qualitative form of the distortions is consistent between the analytical and simulated antimerons, so we can be confident in our conclusions regarding tuning the antimeron properties by varying the micromagnetic parameters. It is also evident, however, that the analytical model underestimates the distortion present in the antimeron simulations, due to the choice of the linear antimeron ansatz. Repeating the calculation for an alternative ansatz that more accurately reproduces the antimeron profile could give a better agreement with the simulation data.

\end{document}